\documentclass[a4paper,11pt,preprintnumbers]{article}
\pdfoutput=1

\usepackage{tikz-feynman}
\usepackage{physics}
\usepackage{makecell}
\usepackage{amssymb,mathrsfs,amsmath}
\usepackage{a4wide}
\usepackage{color}
\usepackage{slashed,soul}
\usepackage{graphicx}
\usepackage{amsfonts}
\usepackage{lscape}
\def\linkcolor{cyan!70!black}

\usepackage[
colorlinks=true
,urlcolor=\linkcolor
,anchorcolor=\linkcolor
,citecolor=\linkcolor
,filecolor=\linkcolor
,linkcolor=\linkcolor
,menucolor=\linkcolor
,linktocpage=true
,pdfproducer=medialab
,pdfa=true
]{hyperref}
\usepackage{amsthm}
\usepackage{booktabs}
\usepackage{array}
\usepackage{rotating}
\usepackage[numbers, sort&compress]{natbib}
\usepackage{multirow}
\usepackage{float}
\usepackage[utf8]{inputenc}
\usepackage[T1]{fontenc}
\usepackage{appendix}
\usepackage{epsfig}
\usepackage{orcidlink}
\usepackage{comment}
\usepackage{adjustbox}
\usepackage{tcolorbox}
\usepackage{booktabs}

\newcommand{\beq}{\begin{equation}} 
\newcommand{\eeq}{\end{equation}} 
\newcommand{\ba}{\begin{array}}  
\newcommand{\ea}{\end{array}} 
\newcommand{\bea}{\begin{eqnarray}}  
\newcommand{\eea}{\end{eqnarray} }  
\newcommand{\bal}{\begin{align}}
\newcommand{\eal}{\end{align}}   
\newcommand{\bi}{\begin{itemize}}  
\newcommand{\ei}{\end{itemize}}  
\newcommand{\ben}{\begin{enumerate}}  
\newcommand{\een}{\end{enumerate}}  
\newcommand{\bc}{\begin{center}}
\newcommand{\ec}{\end{center}} 
\newcommand{\bt}{\begin{table}}
\newcommand{\et}{\end{table}}  
\newcommand{\btb}{\begin{tabular}}
\newcommand{\etb}{\end{tabular}}

\newcommand{\stau}{\tilde{\tau}}

\allowdisplaybreaks

\let\OLDthebibliography\thebibliography
\renewcommand\thebibliography[1]{
  \OLDthebibliography{#1}
  \setlength{\parskip}{0pt}
  \setlength{\itemsep}{0pt plus 0.3ex}
}

\begin{document}

\vspace{1cm}

\begin{titlepage}

\begin{flushright}
 \end{flushright}
\vspace{0.2truecm}

\begin{center}
\renewcommand{\baselinestretch}{1.8}\normalsize
\boldmath
{\LARGE\textbf{
Asymmetric Dark Matter in SUSY with approximate $R-$symmetry
}}
\unboldmath
\end{center}

\vspace{0.4truecm}

\renewcommand*{\thefootnote}{\fnsymbol{footnote}}

\begin{center}

{
Marco Ardu$^1$\footnote{\href{mailto:marco.ardu@ific.uv.es}{marco.ardu@ific.uv.es}},
Daniel Queiroz Correa$^1$\footnote{\href{mailto:daniel.queiroz@uv.es}{daniel.queiroz@uv.es}}
and Oscar Vives$^1$\footnote{\href{mailto:oscar.vives@uv.es}{oscar.vives@uv.es}}
}

\vspace{0.7truecm}

{\footnotesize
$^1$ Instituto de F\'{\i}sica Corpuscular, Universidad de Valencia and CSIC\\ 
 Edificio Institutos Investigaci\'on, Catedr\'atico Jos\'e Beltr\'an 2, 46980 Spain
 \\
}
\vspace*{2mm}
\end{center}

\renewcommand*{\thefootnote}{\arabic{footnote}}
\setcounter{footnote}{0}

\begin{abstract}
We implement the asymmetric dark matter framework, linking the ordinary and dark matter abundances, within a supersymmetric context. We consider a supersymmetric model that respects an approximate $U(1)_R$ symmetry, which is broken in such a way that at high temperature the $R$ breaking sector mediate processes in equilibrium, but at the SUSY mass scale, the sparticles asymmetry is frozen. In this framework, the gravitino serves as the dark matter candidate, and its mass is predicted to be $\sim10$ GeV to match the observed relic abundance. We identify several realistic spectra; however, the requirement for the Next-to-Lightest Supersymmetric Particle (NLSP) to decay into the gravitino before Big Bang Nucleosynthesis constrains the viable spectrum to masses above 2 TeV.
\end{abstract}

\end{titlepage}


\section{Introduction}\label{sec:Intro}
Despite the enormous success of the Standard Model (SM) of the strong and electroweak interactions in describing practically all phenomena observed at high-energy colliders, we remain convinced that the SM is not the ultimate theory. Several reasons support this belief. From the theoretical side: the hierarchy problem, the multitude of free parameters of the model, especially in the flavour sector; the apparently arbitrary choice of symmetries and particle representations; the origin of $CP$ violation; and, last but not least, the unification with quantum gravity.
               
On the experimental front, while the SM is able to explain all the observed phenomena up to energies of a few TeV (with the expection of neutrino masses), the situation changes when we turn to astrophysics and cosmology.  The evidence for the existence of Dark Matter \cite{Zwicky:1933gu,Rubin:1970zza}, coming from a wide range of observations \cite{Bertone:2004pz,Jungman:1995df, Cirelli:2024ssz}, is overwhelming, yet the SM lacks any component that can account for it. Similarly, the Universe is composed almost entirely of matter \cite{Planck:2018vyg}, with no detectable presence of antimatter in the observable, causally connected regions. The SM cannot explain this matter-antimatter asymmetry, as it cannot be dynamically generated within the model \cite{Gavela:1993ts, Huet:1994jb}. 

All these reasons motivate the search for extensions of the SM, ideally a single framework that can simultaneously address all these unresolved questions. A more modest goal, however, would be to identify connections between different observations that could guide us toward a first step in achieving this ambitious objective.

The remarkably similar values of the dark matter and ordinary matter energy densities in the Lambda Cold Dark Matter model ($\Lambda$CDM)  are really surprising, especially considering they are typically assumed to arise from two completely unrelated mechanisms. At the present epoch, dark matter constitutes about $\Omega_{\rm DM}\sim$ 26.5 \%, while  ordinary matter accounts for a  $\Omega_{b} \sim$ 4.9 \% of the energy density of the Universe \cite{Planck:2018vyg}, just a factor of 5 difference between two quantities that could, in principle, differ by many orders of magnitude. This similarity is part of the so-called ``coincidence problem'', that usually refers to the comparable densities of dark energy, dark matter and visible matter. This ``dark matter-visible matter coincidence'' suggests, precisely, the possibility of an associate production of both components of the Universe energy density by a single mechanism.

This idea has been explored in the literature under the name of Asymmetric dark matter models: an asymmetry in the number densities of particles and antiparticles is shared between ordinary and dark matter \cite{Kaplan:1991ah, Nussinov:1985xr, Nardi:2008ix, Kaplan:2009ag,Petraki:2013wwa,Zurek:2013wia}. After the annihilation of the symmetric component, the number densities of baryon and dark matter are determined by the asymmetry, and thus are of the same order. Generally, this connection between the asymmetries is achieved through the introduction of a new dark sector with an independent gauge group and particle content coupled with the SM via a feeble portal interaction \cite{Shelton:2010ta, Haba:2010bm, Frandsen:2011kt}.

As mentioned earlier, this framework can simultaneously account for the baryon and dark matter abundances, effectively addressing the coincidence problem. However, one might argue that this solution comes at the expense of introducing a 
dark sector that serves no other purpose. This raises the question: is it possible to achieve the same result within a more complete (and well-motivated) theory that addresses multiple issues at once?

The Minimal Supersymmetric Standard Model (MSSM) was, and still remains, one of the best example of a model able to solve a wide range of problems simultaneously \cite{Haber:1984rc,Nilles:1983ge,Chung:2003fi}.  It addresses the hierarchy problem, aids in gauge unification, provides a natural mechanism for electroweak symmetry breaking, and even includes candidates for dark matter \cite{Jungman:1995df}. The question then is: could supersymmetry also explain the similarity between the dark matter and visible matter densities?

In a supersymmetric framework, an asymmetry in $B-L$ would be shared by SM fermions and their supersymmetric partners. Additionally, if $R$-parity is conserved, the dark matter abundance is 
given by the density of the
lightest supersymmetric particle (LSP). This leads us to consider a scenario in which the $B-L$ asymmetry could set the abundance of the LSP, naturally solving the dark--visible matter coincidence problem.  
As we will see, although this idea is very simple, it is challenging to implement in the MSSM. First, the LSP must carry a conserved charge to preserve the $B-L$ asymmetry. Second, the model interactions must  efficiently eliminate the symmetric LSP abundance while preserving the asymmetric component. This is challenging in the presence of interactions that can transfer the asymmetry between the scalar and fermion sectors, as happens in the MSSM.  Finally, achieving similar baryon and dark matter abundances would require an LSP mass in the tens of GeV range, which is difficult to reconcile with current experimental constraints. 

In this work, we explore strategies to address these challenges and develop a supersymmetric model capable of implementing the asymmetric dark matter framework. Previous attempts, such as those in Refs.~\cite{Kang:2011ny,DEramo:2011dhr}, extended the MSSM with meta-stable vector-like particles to store the asymmetry, which subsequently decay into the lightest supersymmetric particle (either a gravitino or a bino). Here, we instead focus on the case of an approximate $R$-symmetry, which can preserve the asymmetry stored in a supersymmetric particle of the MSSM serving as the next-to-LSP (NLSP). The NLSP ultimately decays into the gravitino LSP, and its abundance willl be determined by the transferred asymmetry. While this scenario is briefly mentioned in \cite{Kang:2011ny}, it has not been explored in detail. Additionally, we demonstrate that $R$-breaking is essential to establish the connection between the baryon asymmetry and the dark matter abundance.

The paper is organised as follows. In section \ref{sec:Cosmo}, we will study the evolution of a baryon/lepton asymmetry generated at high scales in a supersymmetric model at different temperatures and how to ensure the asymmetry in the supersymmetric sector is preserved thanks to an approximate $R$-symmetry. Section \ref{sec:Models} is devoted to the explicit construction of supersymmetric models that can successfully implement this idea. Section \ref{sec:detcosmo} presents the cosmology of this model and section \ref{sec:pheno} the low-energy phenomenology.

\section{Asymmetry cosmological evolution in Supersymmetry}\label{sec:Cosmo}

As explained in the introduction, the main goal of this work is to explore the conditions required to implement the Asymmetric Dark Matter idea in a Supersymmetrized version of the Standard Model.  

To this end, we assume that a lepton asymmetry is generated at high scales in a supersymmetric Standard Model through leptogenesis in any of its variants \cite{Davidson:2008bu,Affleck:1984fy,Pilaftsis:1997jf,Hambye:2000zs,Hambye:2001eu}. A gravitino LSP will act as the final component of dark matter, so it is important to consider the problem of gravitino thermal overproduction. To prevent the overclosure of the universe with a stable gravitino, the reheating temperature after inflation must be kept low enough, $T_R \lesssim 10^6$~GeV \cite{Moroi:1995fs}, so we assume that the lepton asymmetries are generated at lower temperatures.

Then, we have an asymmetry initially produced in any of the three lepton flavours, $B/3 -L_\alpha$, which is subsequently conserved. This asymmetry is redistributed among particles and sparticles via fast equilibrium interactions. 
The particle and sparticle asymmetries are determined in terms of their chemical potentials, $\mu_i$,
\begin{equation}
Y_{\Delta_i} \equiv Y_i - \bar{Y}_i\equiv \frac{n_i - \overline{n_i}}{s} =
\frac{45~ g_i ~\eta_i}{g_* 12\pi^2 } ~ \frac{\mu_i}{T},\label{defnYD}
\end{equation}
with $g_i$ the internal number of degrees of freedom for particle $i$ and $\eta_i = 1, 2$ for fermions and bosons respectively. For a process in equilibrium, the sum of chemical potential in the initial and final states must be equal.  Therefore, all transitions in equilibrium lead to relations between the different chemical potentials.  We consider temperatures above the Supersymmetry breaking scale, larger than all sparticle masses, although low-enough for all the Yukawa interactions to be in equilibrium. This is also the case for the flavour changing interactions in the quark sector, so that quark multiplets of different flavour have the same chemical potential $\mu_q = \mu_{q_i}$, $\mu_d = \mu_{d_i}$, $\mu_u = \mu_{u_i}$. 

In an MSSM scenario, with the minimal particle content, we have the following chemical potentials\footnote{We consider the chemical potentials of right-handed $SU(2)$ singlets rather than considering $f_R^c$},
\begin{eqnarray}
	  \mu_{q},\ \mu_{u},\ \mu_{d},\ \mu_{\ell_\alpha},\ \mu_{e_\alpha},\ \mu_{H_u},\ \mu_{H_d},\ \mu_{\tilde{B}},\ \mu_{\tilde{W}}, \mu_{\tilde{g}},\nonumber\\
	    \mu_{\tilde{q}},\ \mu_{\tilde{u}},\ \mu_{\tilde{d}},\ \mu_{\tilde{\ell}_\alpha},\ \mu_{\tilde{e}_\alpha},\ \mu_{h_u},\ \mu_{h_d}\,,
	   \label{eq:MSSMmu}
\end{eqnarray}
where we have used that the gauge bosons have zero chemical potential. If all gauginos have Majorana masses, their chemical potentials are also zero, $\mu_{\tilde{B}} = \mu_{\tilde{W}} =\mu_{\tilde{g}} = 0$, and, therefore, particles and sparticles have equal chemical potentials when the gaugino interactions are in equilibrium $\mu_{\tilde{f}} - \mu_f = \mu_{\tilde{B}} = 0$ .
Moreover, the $\mu$-term in equilibrium is a Dirac mass term for the higgsinos $h_u,\ h_d$ and implies  $\mu_{h_u} = - \mu_{h_d} $, which also entails $\mu_{H_u} = - \mu_{H_d} = \mu_H$. 

All in all, we have ten chemical potentials, $ \mu_{q},\ \mu_{u},\ \mu_{d},\ \mu_{\ell_\alpha},\ \mu_{e_\alpha},\ \mu_{H}$, that are determined by the five Yukawa interactions, electroweak sphalerons and hypercharge conservation 
\begin{eqnarray}
	\mu_q-\mu_u-\mu_H=0\nonumber\\
	\mu_q-\mu_d+\mu_{H}=0\nonumber\\
	\mu_{\ell_\alpha}-\mu_{e_\alpha}+\mu_{H}=0 \nonumber\\
	3\sum_{i}\mu_q+\sum_{\alpha}\mu_{\ell_\alpha}=0\nonumber\\
-\sum_{\alpha}\left(\mu_{\ell_\alpha}+ \mu_{e_\alpha}\right) + 2 \mu_H + 3 \left( \mu_q + 2 \mu_u -\mu_d\right)=0 .
\end{eqnarray}
and which can be expressed in terms of the three initial $B/3-L_\alpha$ asymmetries generated by the seesaw sector.  
The solution of these equations gives the distribution of these original asymmetries among the different particles and sparticles at temperatures larger than the SUSY mass (see for instance \cite{Khlebnikov:1988sr, Harvey:1990qw, Davidson:2008bu} for the leptogenesis asymmetry redistribution in the MSSM).

When the temperature reaches the SUSY scale, sparticles begin to annihilate and decay into the (N)LSP and SM particles. We always consider the gravitino as the LSP and, in the following discussion we focus on the case of a right-handed stau as the NLSP, but in section \ref{sec:detcosmo} we will also discuss  other NLSP scenarios.
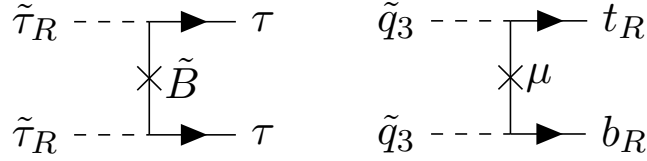
\begin{figure}
	\centering
	\scalebox{1.5}{
		\begin{tikzpicture}
		\begin{feynman}
			\diagram[horizontal=a to b] {
				i1 [particle=\(\tilde{\tau}_R\)] -- [scalar] a -- [fermion] f1 [particle=\(\tau\)],
				i2 [particle=\(\tilde{\tau}_R\)] -- [scalar] b -- [fermion] f2 [particle=\(\tau\)],
				a -- [edge label=\({\tilde{B}}\), insertion=0.5] b,
			};
		\end{feynman}
	\end{tikzpicture}}\qquad 
\scalebox{1.5}{
	\begin{tikzpicture}
		\begin{feynman}
			\diagram[horizontal=a to b] {
				i1 [particle=\(\tilde{q}_3\)] -- [scalar] a -- [fermion] f1 [particle=\(t_R\)],
				i2 [particle=\(\tilde{q}_3\)] -- [scalar] b -- [fermion] f2 [particle=\(b_R\)],
				a -- [edge label=\(\mu\), insertion=0.5] b,
			};
		\end{feynman}
\end{tikzpicture}}
		\caption{bino and Higgsino mediated processes that can remove the sparticles asymmetry. A Majorana bino mass (left-diagram) and $\mu$-term (right-diagram) insertions are necessary to flip the fermion line.}
		\label{fig:asymremov}
\end{figure}
Since the stau only decays to gravitino and tau at late times, for temperatures around the stau mass the stau number density evolves due to the stau annihilations. Right-handed staus annihilate to SM particles, $\tilde \tau_R^- \tilde \tau_R^+ \to {\rm SM}$, removing the symmetric $\tilde \tau_R$ population until the process freezes out. Even if a stau asymmetry survives from the annihilations and decays of the other SUSY particles, in the presence of Majorana gaugino masses, there is a second process, $\tilde \tau_R^- \tilde \tau_R^- \to \tau_R^- \tau_R^-$, proportional to the bino Majorana mass, that will transfer this stau asymmetry to the SM fermions (see the left diagram of Figure \ref{fig:asymremov}). Neglecting the tau Yukawa, the annihilation cross section through a bino exchange is \cite{Asaka:2000zh},
\begin{equation}
\langle \sigma(\stau_R + \stau_R \to \tau \tau) v \rangle
\sim \frac{16 \pi \alpha_{\rm em}^{2} M_{\tilde B}^2} {\cos^4 \theta_W \left(M_{\tilde \tau}^2 + M_{\tilde B}^2\right)^2}~~,~~\label{eq:binowashout}
\end{equation}
where $M_{\tilde B}$ is the Majorana bino mass and $M_{\tilde \tau}$ the stau mass. 
This process decouples when $\Gamma (\stau_R + \stau_R \to \tau \tau) = n_{\tilde \tau_r}^{eq} \times \langle \sigma  v \rangle \simeq H$,
\begin{eqnarray}    
\Gamma (\stau_R + \stau_R \to \tau \tau) & \sim&
\frac{m_{\tilde \tau}^3 }{ z^{3/2} \pi^2}e^{-z}\frac{16 \pi  \alpha_{\rm em}^{2} }{M_{\tilde B}^2 \cos^4 \theta_W}
\approx \frac{  (100)^{1/2} m^2_{\stau}} {z^2 M_{Pl}}\nonumber \\
\Rightarrow {\frac{m_{\tilde \tau}}{T_b}} &\approx &\ln  \left[ 10^{10}\frac { m_{\tilde \tau } }{ \rm TeV } \frac {(10 {\rm TeV})^2 }{  M_{\tilde B}^2 }  \right] \sim
{23 +
\ln  \left[\frac {m_{\tilde \tau }}{ \rm TeV } \frac {(10 {\rm TeV})^2 }{  M_{\tilde B}^2 }\right] } \,,
\end{eqnarray}
where $T_b$ is the decoupling temperature.
This implies that this process freezes out only when the temperature is a factor $\sim 20$ below the stau mass. At those temperatures, the ratio of asymmetry in
the $\tilde \tau_R$ to asymmetry in the $\tau$s will be,
\begin{equation}
\frac{Y_{\tilde\tau} - \overline{Y_{\tilde\tau}}}{Y_\tau - \overline{Y_\tau}}
\approx 
\frac{6\sqrt{\pi/2} z_b^{3/2} e^{-z_b}}{\pi^2}\, \label{eq:washout}
\end{equation}
with $z_b = m_{\tilde \tau}/T_b$.
The surviving asymmetry in $\tilde \tau_R$ is exponentially suppressed, 
$\sim 10^{-10}$ for $z\sim 23$, and would thus be negligible with respect to the baryon asymmetry.

The situation would improve slightly if the bino has a large Dirac mass, $M_{\tilde B_D}$ with a smaller Majorana mass,  $M_{\tilde B_M}$. In this case, the decoupling temperature would go as, 
\begin{eqnarray}    
{\frac{m_{\tilde \tau}}{T_b}} &\approx &\ln  \left[ 10^{10}\frac { m_{\tilde \tau } }{ \rm TeV } \frac {(10 {\rm TeV})^2 }{  M_{\tilde B_D}^2 } \left(\frac {M_{\tilde B_M} }{  M_{\tilde B_D} }  \right)^2 \right] \sim
{23 +
\ln  \left[\frac {m_{\tilde \tau }}{ \rm TeV } \frac {(10 {\rm TeV})^2 }{  M_{\tilde B}^2 }\right] + 2 \ln  \left[\frac {M_{\tilde B_M} }{M_{\tilde B_D} }\right] }\,\label{eq:zbino}
\end{eqnarray}
Then, to get $z \simeq 1$ we would need the Majorana mass to be 10 orders of magnitude smaller than the Dirac bino mass. This means that we need a very small or zero Majorana mass to preserve any relation between the baryon asymmetry and the gravitino (dark matter) number density. 

Moreover, the bino mass is not the only problem: any gaugino Majorana mass would have the same effect, either generating a new contribution to the bino Majorana mass at the loop level or through other asymmetry-removing processes involving particles in chemical equilibrium with the staus. Dangerous transitions also arise from the higgsino $\mu$-term and the trilinear couplings. For example, the co-annihilation of third-generation squark doublets $\tilde{q}_3 \tilde{q}_3 \to t_R b_R$ can be mediated by Higgsino exchange and a $\mu$ insertion, as illustrated in the right diagram of Fig.\ref{fig:asymremov}. Since the top and bottom Yukawa couplings are large, this process freezes out around the same time as the gauge interactions. As a result, the surviving squark doublet asymmetry features an exponential suppression similar to that of Eq.~(\ref{eq:washout}).
If the squark and stau masses are of similar size, as expected in standard SUSY breaking mediation mechanisms, the stau NLSP asymmetry will also be washed out via the fast interactions that maintain chemical equilibrium between the squark doublets and the staus.

These issues can be circumvented by imposing a continuous $R$-symmetry. Under this symmetry, Majorana gaugino masses, trilinear couplings, and the $\mu$-term are forbidden. 

It is clear that an exact $R$-symmetry could preserve the asymmetry in the NLSP. SM particles carry zero $R$-charge, while all supersymmetric particles decay into the NLSP, which becomes the only remaining particle with non-zero $R$-charge at late times. Consequently, a net $R$-charge carried by the sparticles would be transferred to the NLSP asymmetry, which ultimately decays into gravitinos. However, if the $R$-charge is conserved throughout the entire cosmological history, a Universe that starts with no net $R$-charge will lead to a vanishing NLSP asymmetry. Furthermore, even if an initial net $R$-charge is assumed, it would generally be uncorrelated with the lepton asymmetry in minimal leptogenesis scenarios. In either case, the connection between the dark matter abundance and the baryon asymmetry would be lost.

One possibility is to have the $R$-symmetry spontaneously broken at a time such that the phase transition occurs late enough for asymmetry-removing transitions to be already frozen out, but early enough that not all sparticles have decayed into the NLSP. 
However, this scenario is complicated because the spontaneously broken $R$-symmetry would predict a massless Goldstone boson, which would be in serious tension with cosmological observables and laboratory searches. Some explicit breaking is necessary to give the would-be Goldstone a mass, so we focus directly on a scenario where $R$ is only an approximate symmetry, explicitly broken by selected couplings.

The source of the explicit $R$-breaking must be chosen with care. On one hand, we must avoid reintroducing the fast asymmetry-removing transitions that we previously discussed. On the other hand, we need the $R$-breaking couplings to be large enough to mediate processes in equilibrium at sufficiently high temperatures, so that the $R$ charge is not conserved during the whole cosmological evolution. In the next section, we construct a model that satisfies these requirements, and show that can break $R$ charge conservation in such a way that part of the primordial lepton asymmetry can survive in the (N)LSP, providing a natural asymmetric DM candidate. 

\section{$R$-symmetric MSSM}\label{sec:Models}

As discussed in the previous section, to accommodate the asymmetric dark matter idea in supersymmetry, we need an $R$-symmetric model with a controlled soft-breaking. 

$U(1)_{\rm R}$-symmetric models have a long history in the literature \cite{Fayet:1974pd,Hall:1990hq,Randall:1992cq,Nelson:2002ca,Fox:2002bu,Abel:2013kha,Diessner:2014ksa,abel2011easy,benakli2011dirac,benakli2013dirac,benakli2017framework}. The model we build is based in the usual $R$-symmetric superpotential, with $U(1)_{\rm R}$ softly-broken through a new singlet, $\hat S_1$, that interacts only with the MSSM Higgses and has $R$-charge 2. In order to give an $R$-symmetric  mass to the fermion singlet we need a second singlet, $\hat S_2$, with vanishing $R$-charge. The full superpotential reads
\begin{align}
    \nonumber W_{SBR}&=Y_u\, \hat{u}\, \hat{q}\, \hat{H}_u - Y_d\, \hat{d}\, \hat{q}\, \hat{H}_d - Y_e\, \hat{e}\, \hat{l}\, \hat{H}_d + \mu_D\, \hat{R}_d\, \hat{H}_d + \mu_U\, \hat{R}_u\, \hat{H}_u +\\
\nonumber &\lambda_{d}\, \hat{S}\, \hat{R}_d\, \hat{H}_d + \lambda_{u}\, \hat{S}\, \hat{R}_u\, \hat{H}_u + +\Lambda_{d}\, \hat{R}_d\, \hat{T}\, \hat{H}_d + \Lambda_{u}\, \hat{R}_u\, \hat{T}\, \hat{H}_u +  \\ & \tilde \lambda_{d}\, \hat{S_2}\, \hat{R}_d\, \hat{H}_d + \tilde \lambda_{u}\, \hat{S_2}\, \hat{R}_u\, \hat{H}_u + \mu_{S_2}\, \hat{S}_1\, \hat{S}_2 + \mu_{S}\, \hat{S}_1\, \hat{S} + \kappa\, \hat{S}_1^3 + \lambda_1\, \hat{H}_u\, \hat{H}_d\, \hat{S}_1 \label{eq:softly_brkn_Rsymmetric_superpotential}
\end{align}
where the top two lines correspond to the usual $R$-symmetric model. The last line includes all $S_1$ and $S_2$ interactions and the $R$-symmetry breaking term $\kappa \hat{S}_1^3$, while the charges of the different fields are given in Table~\ref{tab:Rcharge}. 

The soft terms for this superpotential must also preserve $R$-symmetry with the only exception of the soft-trilinear term $ B_\kappa S_1^3$ 
\begin{eqnarray} 
- L_{1} &=& \bigg(\frac{1}{2} B_{S} S^{2}+\frac{1}{2} B_T T^{2} +\frac{1}{2} B_{S_2} S_{2}^{2}+  B_{SS_2} SS_{2} + \frac{1}{2} B_O O_{{\alpha}} O_{{\alpha}} +\\
&& \frac{1}{3} B_\kappa S_1^3+\frac{1}{2}B_{S^2 S_2} S^2 S_2+ \frac{1}{2}B_{S_2^2 S} S_2^2 S+\mbox{h.c.} \bigg)+\\
&& m_S^2 |S|^2 + 
 m_{S_1}^2 |S_1|^2 +m_{S_2}^2 |S_2|^2 + m_t^2 |T|^2  +m_O^2|O|^2 \nonumber  \label{W soft}\\ 
- L_{2} &=& m_{H_d}^2 |H_d|^2 +m_{H_u}^2 |H_u|^2 +B_\mu~H_u H_d  +m_{R_d}^2 |R_d|^2 +m_{R_u}^2 |R_u|^2 + \\ 
&&\tilde{d}^*_{L,{i \alpha}} m_{\tilde q,{i j}}^{2} \tilde{d}_{L,{j \alpha}}+\tilde{d}^*_{R,{i \alpha}} m_{\tilde d,{i j}}^{2} \tilde{d}_{R,{j \alpha}} +\tilde{e}^*_{L,{i}} m_{\tilde l,{i j}}^{2} \tilde{e}_{L,{j}} +\tilde{e}^*_{R,{i}} m_{\tilde e,{i j}}^{2} \tilde{e}_{R,{j}}+ \nonumber \\ 
&&\tilde{u}^*_{L,{i \alpha}} m_{\tilde q,{i j}}^{2} \tilde{u}_{L,{j \alpha}}+ \tilde{u}^*_{R,{i \alpha}} m_{\tilde u,{i j}}^{2} \tilde{u}_{R,{j \alpha}} + \tilde{\nu}^*_{L,{i}} m_{\tilde l,{i j}}^{2} \tilde{\nu}_{L,{j}} \nonumber
\\ - L_{3} &=& 2 M^{B}_D \tilde{B} \tilde{S} +  2 M^{S_2}_D \tilde{B} \tilde{S_2} + 
M^{O}_D \tilde{G}_{{\alpha}} \tilde{O}_{{\alpha}}+ 2 M^{W}_D \tilde{T}\tilde{W}
\end{eqnarray} 
 where $L_1$ includes the adjoint scalar  soft terms, $L_2$  the non-adjoint scalars and $L_3$ the Dirac gaugino masses. As can be seen in Table~\ref{tab:Rcharge}, the two singlet superfields $\tilde S$ and $\tilde S_2$ have exactly the same charges and we could think that only one of them, $\tilde S$, could simultaneously give masses to $\tilde S_1$ and $\tilde B$. However, it is clear that we need four fields to provide Dirac masses to the fermionic components. Hence, we need both $\tilde S$ and $\tilde S_2$. $\tilde S$ is chosen to be the combination providing the Dirac mass to $\tilde B$ but the coupling $\tilde S_2$-$\tilde B$, given that it is allowed, is also present in $L_3$. Moreover, we have included a $B_\mu$ term in $L_2$, which is also allowed by all symmetries, including the $R$-symmetry, even though the $\mu$-term is absent in the superpotential.

\begin{table}
\centering
\begin{tabular}{ccc} 
Superfield & $R$-charge & $U(1)_Y \otimes\, SU(2)_L \otimes\, SU(3)_C$ \\ 
\hline \noalign{\vskip 0.1cm} 
$\hat{q}$ & 1 & $(\frac{1}{6},{\bf 2},{\bf 3}) $ \\ 
$\hat{l}$ &  1 & $(-\frac{1}{2},{\bf 2},{\bf 1}) $ \\ 
$\hat{H}_d$ & 0 & $(-\frac{1}{2},{\bf 2},{\bf 1}) $ \\ 
$\hat{H}_u$ &  0 & $(\frac{1}{2},{\bf 2},{\bf 1}) $ \\ 
$\hat{d}$ & 1 & $(\frac{1}{3},{\bf 1},{\bf \overline{3}}) $ \\ 
$\hat{u}$ &  1 & $(-\frac{2}{3},{\bf 1},{\bf \overline{3}}) $ \\ 
$\hat{e}$ & 1 & $(1,{\bf 1},{\bf 1}) $ \\ \hline \noalign{\vskip 0.1cm} 
$\hat{S}$ & 0 & $(0,{\bf 1},{\bf 1}) $ \\ 
$\hat{T}$ & 0 & $(0,{\bf 3},{\bf 1}) $ \\ 
$\hat{O}$ & 0 & $(0,{\bf 1},{\bf 8}) $ \\ 
$\hat{R}_d$ &  2 & $(\frac{1}{2},{\bf 2},{\bf 1}) $ \\ 
$\hat{R}_u$ &  2 & $(-\frac{1}{2},{\bf 2},{\bf 1}) $ \\ \hline \noalign{\vskip 0.1cm}
$\hat{S}_1$ & 2 & $(0,{\bf 1},{\bf 1}) $ \\ 
$\hat{S}_2$ & 0 & $(0,{\bf 1},{\bf 1}) $ 
\end{tabular}
\caption{ $R$-charges of the chiral superfields of the MSSM, the extra fields added to obtain Dirac masses for Higgs and gauginos and the new particles introduced to softly break the R-symmetry.}
\label{tab:Rcharge}
\end{table}

From this superpotential and the soft Lagrangian, we see that the model preserves the $U(1)_{R}$ symmetry with the only exception of the term $\kappa \hat{S}_1^3$ in the superpotential and the soft term $B_\kappa S_1^3$. These terms break $R$-charge by 4 and 6 units, respectively. Then, $\hat S_1$ couples only to $\hat H_{u,d}$ and to $\tilde S$ or $\tilde S_2$ through the superpotential. The main question now is whether these breaking terms can generate gaugino masses or a $R$-breaking mass for the higgsinos, which contribute to the processes identified in Section \ref{sec:Cosmo} and can wash out the sparticle asymmetries. To induce gaugino Majorana masses or the $\mu$ term, which have a spurion $R$ charge of 2, insertions of both couplings are required. The possible combinations of $R$-breaking couplings from the Lagrangian would be: i) $B_\kappa \lambda_1 \kappa^*$, coupling the three scalars $S_1$, $H_d$, $H_u$ at one-loop, ii) $B_\kappa \mu_{S,S_2} \kappa^*$, one loop coupling of the scalar $S_1$ to $S$ or $S_2$ and iii)  $B_\kappa  \kappa^*$ coupling at tree-level two fermionic $\tilde S_1$ to two scalar $S_1$. In Section \ref{sec:detcosmo}, we discuss in more details the radiative contributions of these couplings to the asymmetry wash-out transitions.

\section{Cosmology}\label{sec:detcosmo}
In this section, we discuss the cosmology and the asymmetric DM abundance predicted by the model presented in Section \ref{sec:Models}.
We will show that processes in the $R$-breaking sector remain in equilibrium at high temperatures, generating a non-zero $R$ charge associated with the lepton asymmetries produced by the decay of a heavy seesaw sector. As the temperature drops below the sparticle mass scale, the $R$-breaking transitions responsible for transferring the sparticle asymmetry to the Standard Model freeze out, preserving an NLSP asymmetry that is subsequently transferred to the gravitino LSP.

\subsection{$T\gtrsim m_{\rm SUSY}$}\label{ssec:cosmohighT}
As outlined in Section \ref{sec:Cosmo} for the MSSM, we assume that initial $B/3 - L_\alpha$ asymmetries are generated at high temperatures via the decays of a heavy seesaw sector. These asymmetries are dynamically redistributed among particles and sparticles via fast interactions that enforce chemical equilibrium. In principle, the full set of chemical potentials in the model discussed in Section \ref{sec:Models} should extend the MSSM set (see Eq.(\ref{eq:MSSMmu})) by including the fermion and boson components of the additional superfields:
\begin{align}
	\mu_{S},\ \mu_{\tilde{S}},\ \mu_{T},\ \mu_{\tilde{T}},\ \mu_{O},\ \mu_{\tilde{O}},\ \mu_{R_u},\ \mu_{r_u},\ \mu_{R_d},\ \mu_{r_d},\nonumber\\
	\mu_{S_1},\ \mu_{\tilde{S}_1},\ \mu_{S_2},\ \mu_{\tilde{S}_2}.
\end{align}
Since the gauginos now have Dirac masses, they can in principle possess non-zero chemical potentials which must be equal and opposite to the respective fermion partner, i.e. $\mu_{\tilde{B}} = -\mu_{\tilde{S}}$, $\mu_{\tilde{W}} = -\mu_{\tilde{T}}$, and $\mu_{\tilde{g}} = -\mu_{\tilde{O}}$. Similarly, the $R$-symmetric Higgsino mass terms, $\mu_U$ and $\mu_D$, impose the relations $\mu_{h_u} = -\mu_{r_u}$ and $\mu_{h_d} = -\mu_{r_d}$, which can be inferred from scatterings such as $h_{u,d} X \leftrightarrow r^*_{u,d} X$ featuring  $\mu_{U,D}$ insertions. The Yukawa interactions for quarks in equilibrium impose the following conditions:
\begin{eqnarray}
	\mu_{H_u} - \mu_q + \mu_u &=& 0\nonumber\\
	\mu_{H_d} - \mu_q + \mu_d &=& 0, \label{eq:Yukqtext}
\end{eqnarray}
while fast QCD sphaleron transitions give\footnote{The octet and gluino chemical potentials cancel in the sum}:
\begin{eqnarray}
	2\mu_q - \mu_u - \mu_d = 0. \label{eq:QCDsph}
\end{eqnarray}
Summing Eqs.~(\ref{eq:Yukqtext}) and (\ref{eq:QCDsph}), we find that the scalar Higgs up and down chemical potentials sum to zero: $\mu_{H_u} + \mu_{H_d} = 0$. Considering sufficiently large $R$-breaking soft term $B_\kappa S^3_1$ and Yukawa coupling $\lambda_1 h_u h_d S_1$ such they mediate the processes in equilibrium, we also find that the fermion components of the Higgs superfields have opposite chemical potentials:
\begin{equation}
	\mu_{h_u} + \mu_{h_d} = \mu_{S_1} = 0.
\end{equation}
This implies that, as in the MSSM, the gaugino chemical potentials vanish, meaning that particles and sparticles share the same chemical potential (see Appendix \ref{app:chemeqs} for details). By combining all relevant interactions and processes, the asymmetries can ultimately be expressed in terms of the initial $B/3 - L_\alpha$ asymmetries, denoted as $Y_{\Delta_\alpha}$. This gives:
\begin{equation}
	\Delta \hat{Y}_i = c_{i\alpha} Y_{\Delta_\alpha},\label{eq:highTasym}
\end{equation}
where $i$ runs over all particles and sparticles in the model, and $\alpha = e, \mu, \tau$. The values of the $c_{i\alpha}$ coefficients, derived from the chemical equilibrium equations, are provided in Appendix \ref{app:chemeqs}.

\subsection{$T\lesssim m_{\rm SUSY}$ and NLSP decay}\label{ssec:lowT}
After crossing the SUSY breaking scale, the sparticles will start to decay and annihilate into the SM particles and the NLSP. Given that we consider NLSP masses of $m_{\rm NLSP} \gtrsim 1$ TeV, the symmetric relic density left after the NLSP annihilations freeze-out could be sizable \cite{Hisano:2006nn}. However, the NLSP will eventually decay into the gravitino LSP. If the asymmetric contribution to the gravitino density accounts for the observed DM abundance, the gravitino must have a mass $m_{3/2} \lesssim 10$ GeV. Thus, even if the symmetric NLSP abundance is of the same of order as the observed DM density, i.e., $\Omega_{\rm NLSP} \sim \Omega_{\rm DM}$, the contribution to the gravitino abundance would be $ (m_{3/2}/m_{\rm NLSP}) \Omega_{\rm NLSP} \ll \Omega_{\rm DM}$. Therefore, we assume that the NLSP symmetric contribution to the DM abundance is negligible, and we focus exclusively on the evolution of the asymmetry.

\begin{figure}[t]
\centering
\begin{tikzpicture}[scale=2]
	\begin{feynman}
		\vertex (a1) at (-2,0) {\(\tilde{B}\)};
		\vertex  (a2) at (-1,0);
		\vertex  (a3) at (0,0);
		\vertex (a4) at (1,0);
		\vertex (a5) at (2,0) {\(\tilde{B}\)};
		\vertex[dot, label=\({B_\kappa}\)] (b) at (0,0.7) {};
		\vertex[dot, label=\({\lambda_1 \kappa^*}\)] (c) at (0,1.2) {};
		\diagram* {
			(a1) -- [fermion] (a2),
			(a2) -- [anti fermion, edge label'=\({h_u}\)] (a3),
			(a3) -- [fermion, , edge label'=\({h_d}\)] (a4),
			(a4) -- [anti fermion] (a5),
			(a2) -- [charged scalar, quarter left, edge label=\({H_u}\)] (c),
			(a4) -- [charged scalar, quarter right, edge label'=\({H_d}\)] (c),
			(a3) -- [charged scalar,  edge label'=\({S_1}\)] (b),
			(b) -- [anti charged scalar, half left] (c),
			(b) -- [anti charged scalar, half right] (c),
		};
	\end{feynman}
\end{tikzpicture}\qquad
\begin{tikzpicture}[scale=2]
	\begin{feynman}
		\vertex  (a2) at (-1,0);
		\vertex  (a3) at (0,0);
		\vertex (a4) at (1,0);
		\vertex[dot] (b) at (0,0.5) {};
		\vertex[dot] (c) at (0,1) {};
        \vertex[dot, label=\({B_\mu}\)] (d) at (0,1.5) {};
		\diagram* {
			(a2) -- [anti fermion, edge label'=\({h_u}\)] (a3),
			(a3) -- [fermion, , edge label'=\({h_d}\)] (a4),
			(a3) -- [charged scalar,  edge label'=\({S_1}\)] (b),
			(b) -- [anti charged scalar, half left, edge label=\({S_1}\)] (c),
			(b) -- [anti charged scalar, half right, edge label'=\({S_1}\)] (c),
            (c) -- [charged scalar, half left, edge label=\({H_u}\)] (d),
			(c) -- [charged scalar, half right, edge label'=\({H_d}\)] (d),
		};
	\end{feynman}
\end{tikzpicture}
\caption{Three loop contribution to the bino Majorana mass and two-loop contribution to the $\mu$ higgsino mass ($\mu \tilde h_u \tilde h_d$) from the $R$-breaking couplings discussed in the text. }
\label{fig:Rbreaking}
\end{figure}
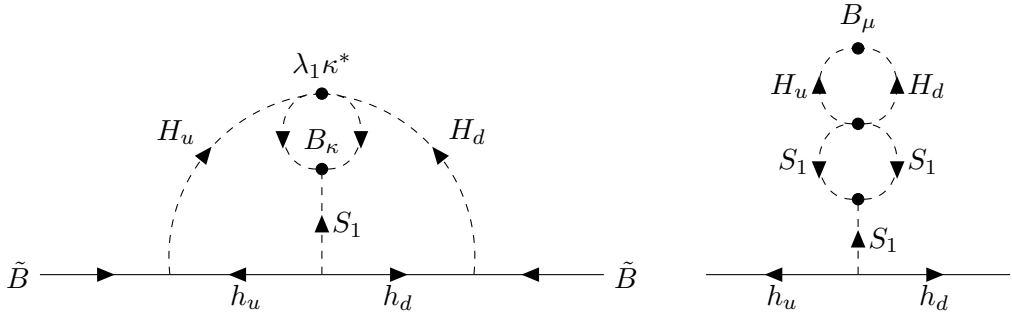

In Section \ref{sec:Cosmo}, we identified some processes (see for instance Figure \ref{fig:asymremov}) that can transfer the sparticle asymmetries to the SM, which are mediated by interactions that violate the $R$-symmetry, which include gaugino Majorana masses or the higgsino $h_u h_d$ mass. As discussed at the end of Section \ref{sec:Models}, the model under consideration generates these $R$-breaking parameters with coupling combinations that must have a spurion charge 2. To generate a gaugino Majorana mass, the $R$-breaking couplings must be inserted into a gaugino line, for instance $\tilde B$. We find that, when the electroweak symmetry is unbroken, the only way to connect these $R$-symmetry breaking terms to $\tilde B$ is through the Higgs fields at three or higher loops. An example of this is illustrated in Figure \ref{fig:Rbreaking}, where a bino Majorana mass is generated via insertions of $\kappa$ and $B_\kappa$. We estimate the size of this three-loop induced bino Majorana mass as
\begin{equation}
	M_1 \sim \left(\frac{1}{16\pi^2}\right)^3 \lambda_1^2 g_1^2 \kappa B_\kappa. \label{eq:binoRbreak}
\end{equation}
Even assuming $B_\kappa$ is of the same order as the gaugino Dirac masses, it is sufficient to take $\lambda_1^2 \kappa \lesssim 10^{-3}$ in order to achieve a Majorana bino mass of $M_1 \sim 10^{-10} M^B_D$. As shown in Eq.~(\ref{eq:zbino}), this is enough to avoid the exponential suppression of the asymmetry. 
Regarding the higgsino mass, $\mu h_u h_d$, which can mediate the asymmetry removing stop-stop annihilation discussed in Section \ref{sec:Cosmo}, the $R$-breaking sector can induce it at two or higher loops (an example is shown in the right-diagram of Figure \ref{fig:Rbreaking}). The estimated higgsino mass is
\begin{equation}
    \mu\sim \left(\frac{1}{16\pi^2}\right)^2 \lambda_1 \kappa B_\kappa \label{eq:muRbreak}
\end{equation}
If we impose a requirement on $\mu$ similar to the one of the Majorana bino mass, that is $\mu\lesssim 10^{-10}\times m_{\rm SUSY}$, we need $\lambda_1 \kappa B_\kappa\lesssim 10^{-6} \times m_{\rm SUSY}$. This requirement is more stringent than necessary because the dangerous processes featuring $\mu$ insertions typically involve stops, which is not the NLSP in the scenarios of interest. If there is a small hierarchy between the stop mass and the NLSP mass, the Boltzmann suppression of the NLSP-to-SM asymmetry can be milder. This is because if the $\mu$ mediated processes transfering the stops asymmetry freeze out at temperatures $m_{\tilde{t}}/T\sim z_{\tilde{t}}$, the Boltzmann suppression for the NLSP asymmetry would be $\exp(-z_{\tilde{t}} \times (m_{\rm NLSP}/m_{\tilde{t}}))$. Even with a mild hiearchy between stops and the NLSP, requiring a sizable surviving asymmetry can be satified for larger $\lambda_1 \kappa B_\kappa$\footnote{For instance, if $m_{\tilde{t}}\sim 10\times m_{\rm NLSP}$ and $m_{\rm NLSP}\sim$ TeV, then $\mu\sim 10^{-5}\times m_{\rm SUSY}$ is sufficient to have an unsuppressed NLSP asymmetry (these estimates follow from calculations similar to those presented in Eq.~(\ref{eq:zbino}))}. 

The above estimates concern the $T=0$ contributions to the $R$ breaking parameters. However, since the relevant temperature for the asymmetry removing transitions is $T\sim m_{\rm SUSY}$, thermal effects can be important. In particular, the loop suppression pre-factors featuring in Eqs (\ref{eq:binoRbreak}) and (\ref{eq:muRbreak}) may not be present in the $T\neq 0$ contributions. Physically, this can be interpreted as a correction to two-point functions arising from a Bino or pair of Higgsinos scattering off with on-shell particles present in the thermal bath, effectively corresponding to a "cut" in the $T=0$ diagrams of Fig.~\ref{fig:Rbreaking}. Although the thermal contributions pre-factors are difficult to estimate, especially for two and three loop diagrams, thermal masses should not change the properties of the bath and should not break global symmetries. Therefore, we expect the thermal effects to feature the same spurion combinations of couplings previously identified from the $T=0$ diagrams. Assuming only one scale $T\sim B_\kappa\sim m_{\rm SUSY}$ for all sparticle masses involved, we expect the thermal contributions to be:
\begin{equation}
    M_1(T)\propto \lambda^2_1 \kappa^* T\qquad \mu(T)\propto \lambda_1 \kappa T \label{eq:thermalRBreaking}
\end{equation}
On the other hand, if the $R$ breaking singlet $S_1$ is heavier than the rest of the SUSY particles, its number density in the thermal bath at temperatures $T\sim m_{\rm SUSY}$ is Boltzmann suppressed. As a result, the thermal masses of Eq.~(\ref{eq:thermalRBreaking}) are either Boltzmann suppressed or include a $\sim 1/(16\pi^2) B_\kappa/M^2_{S_1}$ factor, which arises from integrating out the $S_1$ in the diagrams of Fig.~\ref{fig:Rbreaking}. 

Since the $\kappa$ coupling does not enter in the equilibrium transitions considered in Section \ref{ssec:cosmohighT}, we can control the size of the thermal contribution of Eq.~(\ref{eq:thermalRBreaking}) by taking a sufficiently small $\kappa$ and/or by considering heavier $S_1$ as discussed above. Numerically, we consider values such that the estimated gaugino masses or $R$ breaking higgsinos mass are $\lesssim 10^{-10}\times m_{\rm SUSY}$\footnote{In the parameter space points considered in Section \ref{sec:pheno}, we take $\lambda_1 \sim 10^{-3}$ and $\kappa \sim 10^{-6}$, ensuring that the product entering in thermal masses is $\lambda_1 \kappa \sim 10^{-9}$. Additionally, $S_1$ is approximately ten times heavier than the NLSP, so the Boltzmann or loop factor further suppresses the thermal effects.
}, so that, as can be seen from Eq.~(\ref{eq:zbino}), the decoupling temperatures for the processes that wash-out the sparticles asymmetry is $T\sim m_{\rm SUSY}$.  
As a result, the total sparticle asymmetry  which arose from the high-temperature equilibrium interactions remain constant once the temperatures drops below the sparticle mass scale, while the $R$-conserving annihilations remove the symmetric component of the sparticle densities. Eventually, the sparticles decay, transferring their asymmetries to the NLSP:
\begin{equation}
	\Delta Y_{\rm NLSP} = \Delta \hat{Y}_{\rm NLSP} + \frac{1}{R_{\rm NLSP}} \sum_{A} R_A \Delta \hat{Y}_{A} \label{eq:Rcharge}
\end{equation}
where $R_{\rm NLSP}$ is the $R$ charge of the NLSP, the hatted quantities correspond to the asymmetries found in Eq.~(\ref{eq:highTasym}), and the sum over $A$ includes all $R$-charged particles with charge $R_A$.
\begin{table}[t]
	\centering
\begin{tabular}{|c|c|c|}
	\hline
	NLSP & $m_{3/2}$ & BBN bound on $m_{\rm NLSP}$\\
	\hline
	\hline
	$\tilde{\ell}_R$ & $8.3$ GeV & $ \gtrsim 2$ TeV \\
	\hline
	$\tilde{\ell}_L$ & $5.3$ GeV & $ \gtrsim 2$ TeV \\
	\hline
	$\tilde{\nu}_\alpha$ & $5.3$ GeV & $\gtrsim 2$ TeV \\
	\hline
\end{tabular}
\caption{Predicted gravitino mass ($m_{3/2}$) under the assumption that no asymmetry washout occurs after the temperature crosses the SUSY breaking scale, for different NLSP scenarios. The gravitino mass is determined using Eq.~(\ref{eq:gravitinomass}), by enforcing that the gravitino abundance matches the observed Dark Matter density. The third column provides the lower bound on the NLSP mass, ensuring a lifetime consistent with BBN predictions on primordial element abundances. The lightest neutralino as NLSP is highly constrained by BBN \cite{Feng:2004zu}, so we do not consider this scenario.  }
\label{tab:gravitinomass}
\end{table}
After the heavy sparticles have decayed, the SM asymmetries are still changing according to the equilibrium interactions while the NLSP relic asymmetric density is frozen. The chemical potential equations before electroweak sphalerons go out of equilibrium are given by:
\begin{itemize}
\item Yukawa interactions
\begin{eqnarray}
	\mu_q-\mu_u-\mu_H=0\nonumber\\
	\mu_q-\mu_d+\mu_{H}=0\nonumber\\
	\mu_{\ell_\alpha}-\mu_{e_\alpha}+\mu_{H}=0,\label{eq:chempotlowenergy1}
\end{eqnarray}
\item Electroweak sphalerons 
\begin{eqnarray}
	3\sum_{i}\mu_q+\sum_{\alpha}\mu_{\ell_\alpha}=0,
\end{eqnarray}
\item Hypercharge conservation
\begin{equation}
	\frac{T^2}{6s}\left(-\sum_{\alpha}(\mu_{\ell_\alpha}+\mu_{e_\alpha})+4\mu_H+3(\mu_q+2\mu_u-\mu_d)\right)+\mathcal{Y}_{\rm NLSP} \Delta Y_{\rm NLSP}=0 \label{eq:chempotlowenergy2}
\end{equation}
\end{itemize}
where $\mathcal{Y}_{\rm NLSP}$ is the NLSP hypercharge. Throughout the whole cosmogical hystory the $B/3-L_\alpha$  asymmetries are conserved, and thus we can impose these initial conditions to the equations system above. Once electroweak sphaleron transitions go out of equilibrium, baryon number-changing processes cease, and the ratio $\Delta Y_{\rm NLSP}/\Delta Y_B$ between the NLSP and the baryon asymmetry is calculated.
Finally, the NLSP will decay to gravitinos and SM particles, transfering the asymmetry to the gravitino. If the NLSP can decay hadronically, this decay must occur before Big Bang Nucleosynthesis (BBN) to avoid disrupting the primordial element abundance predictions. Depending on the NLSP, this places constraints on the possible values of both NLSP and gravitino masses. Within our assumptions, the gravitino mass is predicted once the $\Delta Y_{\rm NLSP}/\Delta Y_B$ ratio is known, because its relic density must match the observed DM abundance
\begin{equation}
		\frac{\Delta Y_{\rm NLSP} m_{3/2}}{\Delta Y_{B} m_p}=\frac{\Omega_{\rm DM}}{\Omega_{\rm b}}\sim 5\to m_{3/2}=5 m_p\times \frac{\Delta Y_B}{\Delta Y_{\rm NLSP}}\label{eq:gravitinomass}
\end{equation}
where $m_p$ is the proton mass. If the $R$-breaking parameters are sufficiently large to keep asymmetry-removing processes in equilibrium after crossing the SUSY scale, the final asymmetry of the NLSP will experience partial washout. This can be expressed taking $\Delta Y_{\rm NLSP}/\Delta Y_B \to \epsilon \Delta Y_{\rm NLSP}/\Delta Y_B$, where $0 < \epsilon \leq 1$, modifying the prediction of Eq.~(\ref{eq:gravitinomass}) accordingly
\begin{equation}
	m_{3/2}=5 m_p\times\frac{1}{\epsilon}\times \frac{\Delta Y_B}{\Delta Y_{\rm NLSP}}
\end{equation}
In general, the value of $\epsilon$ is model-dependent and can be determined by solving the Boltzmann equations for the non-equilibrium dynamics of the sparticles annihilation and decay. In Table \ref{tab:gravitinomass}, we write the predicted gravitino mass assuming a negligible effect of the $R$-breaking sector since, in the models we consider, the $R$-breaking couplings are fixed to be small enough to have a wash-out factor $\epsilon \simeq 1$. Note that the gravitino mass predictions are independent of how the initial lepton asymmetries are redistributed among the different flavours, as both the NLSP asymmetry (related to the net $R$ charge) and the baryon asymmetry are proportional to the sum $\sum_{\alpha} Y_{\Delta_\alpha}$.

In the same table, we also list the BBN constraints on the NLSP mass. For example, in the case of a $\tilde{\tau}_{L,R}$ NLSP, the lifetime is given by the rate of the dominant decay channel $\tilde{\tau} \to \tau + \tilde{G}$ \cite{Boubekeur:2010nt}:
\begin{equation}
	t_{\tilde{\tau}}\sim 1.6\ {\rm sec}\ \left(\frac{2\ {\rm TeV}}{m_{\tilde{\tau}}}\right)^5\left(\frac{m_{3/2}}{10\ {\rm GeV}}\right)^2
\end{equation}
In the case of the $\tilde{\tau} \to \tau + \tilde{G}$ decay, the high-energy taus subsequently decay into leptons or mesons. These mesons, however, decay by producing electromagnetic cascades before engaging in significant hadronic interactions, contributing primarily to electromagnetic (EM) energy release. The photons resulting from these electromagnetic cascades thermalize quickly with the background radiation. For life-times $t \leq 10^4$~sec, the maximum energy of the resulting photons is too low to destroy any light elements, making EM BBN constraints relatively weak. In contrast, hadronic energy release is more restrictive, even for decays with small branching ratios. For instance, the three-body decay $\tilde{\tau} \to \tau Z \tilde{G}$, produces hadronic energy when the $Z$ boson decays hadronically. For a $\tilde{\tau}$ of mass 2~TeV decaying at $t < 10^5$~sec, the hadronic branching ratio is approximately $\sim 2 \times 10^{-3}$ \cite{Feng:2004zu}. These hadronic constraints, for a branching ratio of $10^{-3}$, were calculated in \cite{Kawasaki:2004yh}, where the $^4$He mass fraction for a $\tilde{\tau}$ decaying with a lifetime $t \leq 1$–$10$~sec requires that $m_{\tilde{\tau}} Y_{\tilde{\tau}} \leq 10^{-7}$–$10^{-8}$~GeV. Thus, for $m_{\tilde{\tau}} = 2$~TeV, we need $Y_{\tilde{\tau}} \leq 2 \times 10^{-10}$–$10^{-11}$, which is close but still larger than the value expected from the asymmetry without washout, $\Delta Y_{\tilde{\tau}} \sim 1/2 \times \Delta Y_{B} \sim 4 \times 10^{-11}$. Larger $\tilde{\tau}$ masses would further shorten its lifetime, thereby relaxing the BBN bounds even more. A similar discussion holds for sneutrinos and other slepton, because the branching for the hadronic release $\tilde{\ell}\to (Z/\gamma^*)\ell \tilde{G}\to \bar{q} q \ell \tilde{G}$ is expected to be similar.

\section{Phenomenology}\label{sec:pheno}

In the previous sections, we have defined our model and demonstrated that it can implement the asymmetric dark matter idea within a supersymmetric framework. Next, we will show that it is indeed possible to find a viable spectrum in this model, consistent with all experimental constraints, that successfully relates the dark matter and baryon abundances.

The exploration of the viable parameter space in this model is done with  \texttt{SPheno-4.0.5} using two-loop RGEs and one-loop finite corrections \cite{porod2003spheno,Porod_2012,Pierce:1996zz} (also two-loop corrections to the neutral Higgs boson masses \cite{Degrassi:2001yf,Brignole:2001jy,Brignole:2002bz} and to the mu-parameter \cite{Dedes:2002dy,Dedes:2003km} are included), while the model is defined with the Mathematica package \texttt{SARAH-4.15.2} \cite{staub2012sarah,Staub_2014}. Some details on how to obtain the points can be found in Appendix \ref{ap:superspectr}. 
 
In fact, the model defined in Section~\ref{sec:Models} is essentially an $R$-symmetric Dirac gaugino model, as the $U(1)_R$-breaking terms are confined to a separate singlet sector that has minimal impact on the charged supersymmetric spectrum, with the only exception of the Dirac gaugino masses.  Then, we have to deal with the usual problems of Dirac Gaugino models.  

One of the well-known issues with these models is the appearance of tachyonic states \cite{abel2011easy, benakli2011dirac}, a problem that becomes even more pronounced when attempting to achieve the correct Higgs mass \cite{goodsell2023active}.  This difficulty arises because the Higgs mass is further reduced due to mixing with singlet and triplet states, requiring substantial radiative corrections to compensate these contributions \cite{kotlarski2016analysisrsymmetricsupersymmetricmodels}, which in turn narrows the viable parameter space. As shown in Appendix \ref{ap:superspectr}, we avoid the presence of tachyons through some approximate conditions, $B_\mu <0$, $\lambda_U \lesssim - g_1 M_D^B/(\sqrt{2}~ \mu_U)$, $\Lambda_U \lesssim - g_2 M_D^W/(\sqrt{2}~ \mu_U)$. Even fulfilling these tree-level conditions, loop corrections can produce tachyonic states and the absence of tachyons must be checked numerically with 
\texttt{SPheno-4.0.5}. We check also that we have the correct electroweak symmetry breaking, with no directions unbounded from below or charge and color-breaking minima \cite{Nilles:1982dy,Alvarez-Gaume:1983drc,Claudson:1983et,Gunion:1987qv,Casas:1995pd,Camargo-Molina:2014pwa}. 
 
\begin{table}[]\centering
    \begin{tabular}{cc|cc}
        \hline
        \textbf{Parameter} & \textbf{Value} &\textbf{Parameter} & \textbf{Value}\\
        \hline
        $m_{0}$ & 3.5~TeV  & $\tan \beta$ & 15\\
         $M_D^W$ & 5~TeV &  $M_D^O$ & 6~TeV  \\
         $\mu_d$ & 4~TeV &   $\mu_u$ & 7~TeV \\
         $v_S$ & 0.7~GeV &   $v_T$ & -0.2~GeV \\
        $m_{S_1}^2$ & 500~TeV$^2$ &  $m_{S_2}^2$ & 500~TeV$^2$ \\
       $\lambda_u$ & -0.3 &   $\lambda_d$ & -0.3 \\    $\Lambda_u$ & -0.3 &   $\Lambda_d$ & 0.5 \\   
          $\tilde \lambda_u$ & $10^{-3}$ &   $\tilde \lambda_d$ & $10^{-3}$\\    $\lambda_1$ & $10^{-3}$ &   $\kappa$ & $10^{-6}$ \\   
        $B_O$ & -10~TeV$^2$ & $B_\kappa$ & 2~TeV \\   
        $B_S$ & 0~TeV &  $B_T$ & 0~TeV \\\hline 
        $m_{R_d}^2$ & (300,100) ~TeV$^2$ & $m_{R_u}^2$ & (1,15) ~TeV$^2$ \\
        $B_\mu$ & (1,10)~TeV$^2$ & $\mu_{S_2}$ & (5,7) ~TeV \\
        $M_D^B$ & (5,6.5)~TeV &  &
    \end{tabular}
    \caption{Parameters for the benchmark point with $\tilde \tau_R$ or $\tilde \nu$ NLSPs. The variables in white are common for both scenarios while the variables in grey are different for the two cases. The first value corresponds to stau NLSP ($m_{\tilde{\tau}_R}=2.09$~TeV) and the second to sneutrino NLSP ($m_{\tilde{\nu}}=2.42$~TeV). The value $m_0$ corresponds to the SUSY mass, $m_{\tilde q}^2 = m_{\tilde u}^2 = m_{\tilde d}^2 = m_{\tilde l}^2 = m_{\tilde e}^2 = m_O^2 = m_0^2$, with all matrices universal. We take $B_{S_2}=B_{S^2 S_2} = B_{S S_2^2}=0$. The remaining parameters, $m_{H_u}^2$, $m_{H_d}^2$, $m_{S}^2$, $m_{T}^2$, $B_{S S_2}$ and $M_D^{S_2}$ are fixed by solving the tadpole equations.}\label{tab:param_wp_1}
\end{table}

To find a viable parameter space region, we must also require the NLSP to be heavier than $\sim 2$ TeV to avoid any impact on BBN. Moreover, squarks should be heavier than 1.9 TeV to prevent gluino overproduction \cite{atlas2017search}, and gluinos must be heavier than 2 TeV \cite{chalons2019lhc}. In Tables~\ref{tab:param_wp_1} and \ref{tab:working_points_1} we present two benchmark points in the viable region of the parameter space, with a $\tilde \tau_R$ or a $\tilde \nu$ as the NLSP. 
\begin{table}[]\centering
    \begin{tabular}{cc|cc}
        \hline
         \textbf{Particle} & \textbf{Mass} (TeV) & \textbf{Particle} & \textbf{Mass} (TeV) \\
        \hline
        $(H^0_2, H^0_6, H^0_8)$ & $(3.8, 22.4, 75.9)$ & $(H^0_2, H^0_6, H^0_8)$ & $(7.5,22.9,120.6 )$ \\
         $(A_1,A_5,A_7)$ & $(3.6, 22.4, 75.6)$ & $(A_1,A_5,A_7)$ & $(3.2, 22.9, 120.4)$ \\ 
         $(H^\pm_1, H^\pm_3, H^\pm_5)$ & $(3.6,~7.8,15.1)$ & $(H^\pm_1, H^\pm_3, H^\pm_5)$ & $(3.2, ~8.3, ~12.0)$ \\ 
        $(\chi^0_1,\chi^0_3,\chi^0_{10})$ & $(2.2, 4.2,7.3)$ & $(\chi^0_1,\chi^0_3,\chi^0_{10})$ & $(2.8, 4.4, 9.5)$ \\
        $(\chi^\pm_1,\chi^\pm_2,\chi^\pm_4)$ & $(4.2,4,3,7.3)$ & $(\chi^\pm_1,\chi^\pm_2,\chi^\pm_4)$ & $(4.2,4.4,7.3)$ \\
        $(\tilde{u}_1,\tilde{u}_3,\tilde{u}_6)$ & $(4.6,5.0,6.0)$ & $(\tilde{u}_1,\tilde{u}_3,\tilde{u}_6)$ & $(4.0,4.9, 5.3)$ \\
        $(\tilde{d}_1,\tilde{d}_4,\tilde{d}_6)$ & $(4.5, 4.6,5.0)$ & $(\tilde{d}_1,\tilde{d}_3,\tilde{d}_6)$ & $(4.9,5.2,5.3)$ \\
        $(\tilde{e}_1,\tilde{e}_4,\tilde{e}_6)$ & $(2.1,2.3, 3.9)$ & $(\tilde{e}_1,\tilde{e}_4,\tilde{e}_6)$ & $(2.4,4.4,4.7)$ \\
        $(\tilde{\nu}_1,\tilde{\nu}_2,\tilde{\nu}_3)$ & $(3.9, 3.9,3.9)$ & $(\tilde{\nu}_1,\tilde{\nu}_2,\tilde{\nu}_3)$ & $(2.4, 2.7, 2.7)$ \\
        $\tilde{g}_{1,2}$ & $16.1$ & $\tilde{g}_{1,2}$ & $16.2$ \\
        $\phi_O$ & $29.0$ & $\phi_O$ & $29.0$ \\
        $\sigma_O$ & $5.6$ & $\sigma_O$ & $5.6$ \\
        \hline 
        $m_{\tilde{e}_1}$ & 2.1 & $m_{\tilde{\nu}_1}$ & 2.4  \\ 
        $m_{h}$ & 0.1211 & $m_{h}$ & 0.1211 
    \end{tabular}
    \caption{Benchmark spectrum with $\tilde \tau_R$ (left) or $\tilde \nu$ (right) NLSPs for the parameters in Table~\ref{tab:param_wp_1}. }\label{tab:working_points_1}
\end{table}

The resulting spectra are shown in Table~\ref{tab:working_points_1}. It is important to notice that Dirac gaugino masses do not contribute to sfermion masses through Renormalization Group Equations (RGEs), but only via threshold corrections \cite{Fox:2002bu}. As shown in Appendix~\ref{ap:superspectr}, gauge and Yukawa corrections to sfermion masses are still present in the RGEs, explaining the differences between squark and slepton masses, which originate from a common $m_0$ at the high scale. As is usual in the MSSM, third-generation sfermions are lighter at the electroweak scale due to their large Yukawa couplings. Consequently, the right-handed stau or sneutrino can be the NLSP by adjusting the parameters entering the beta function, as discussed in Appendix \ref{ap:superspectr}.

Dirac gaugino masses also evolve with beta functions proportional to the gauge couplings and trilinear parameters, $\lambda_i$  and $\Lambda_i$. The beta function for the gluino Dirac mass is large and negative, $\beta_{M^O_D}\simeq - 6 g_3^2 M^O_D$, while the bino and wino beta functions are smaller and positive (see Eq.~(\ref{eq:betafunc})). This means that, starting from a common mass at a high scale, at the electroweak scale the gluino mass will be significantly larger than the initial Dirac mass value at the high scale. 

This explains the main features of the spectrum in Table~\ref{tab:working_points_1}. Squarks are above 4.5 TeV, and only sleptons are relatively close to the NLSP mass. The gluino is around 16 TeV, much heavier than the other gauginos. The wino is approximately 4 TeV, as indicated by the lightest chargino, while the bino, the lightest neutralino, is 2.2 or 2.8 TeV in the two scenarios. Higgs scalars and pseudoscalars are heavier than 3.5 TeV. In fact, the $S_1$ singlet, associated with explicit $R$-symmetry breaking, corresponds to $H^0_6$ and $A_5$ that have masses roughly ten times larger than the NLSP, suppressing further the $R$-breaking effects discussed in Section \ref{ssec:lowT}.  The scalar gluon is heavy because its tree-level mass is given by $m_{\phi_O}^2=4M^{D\, 2}_O+B_O
$, while the pseudoscalar gluon is lighter but still above 5 TeV. The overall heaviness of the spectrum is a result of the requirement for a slepton NLSP with a mass above 2 TeV.

The choice of the NLSP (between the stau and sneutrino) is primarily influenced by the values of $m_{R_u}^2$ and $m_{R_d}^2$, as we show in Appendix \ref{ap:superspectr}.  These values also affect the Higgs masses at tree level and other masses through loop effects, which explains the different spectra in both scenarios. Notably, in the sneutrino NLSP case, the mass difference between the sneutrino and the lightest charged slepton is very small.

Given the heavy spectrum obtained in our model, these states can not be produced at the LHC. Current LHC limits on Dirac gaugino models are below 2 TeV for gluinos and 1.5 TeV for squarks \cite{chalons2019lhc}. However, the heavy strongly coupled states in our benchmark scenarios could still be studied at a 100 TeV collider \cite{GrillidiCortona:2016pyt}.

Electroweakinos and sleptons, due to their smaller production cross sections, have LHC bounds below 1 TeV \cite{Goodsell:2020lpx,Carpenter:2021tnq}. In our scenario, the lightest slepton (the NLSP) is long-lived on collider scales and interacts with detectors similarly to muons. These states leave a charged track from ionization energy loss, with small energy deposits in calorimeters. The main background consists of muons, that are  only distinguished by the larger stau mass. The staus’ smaller $\beta$ can be measured using time-of-flight to the outer detectors or ionization energy loss \cite{Feng:2015wqa}. At the LHC, searches for disappearing tracks and heavy stable charged particles by CMS and ATLAS barely reach 1 TeV \cite{ATLAS:2017oal,ATLAS:2022rme,CMS:2018rea,CMS:2020atg,Goodsell:2021iwc}. The high-luminosity LHC  with 3 ab$^{-1}$ is expected to exclude long-lived stau masses up to 1.2~TeV \cite{Feng:2015wqa}. At a 100~TeV collider, long-lived staus could be searched for similarly, with exclusions for right-handed staus up to 3 TeV with 2 ab$^{-1}$ \cite{Feng:2015wqa}.

For a sneutrino NLSP, long-lived particle detectors like MoEDAL-MAPP \cite{MoEDAL:2014ttp,Pinfold:2019zwp}, CODEX-b \cite{Aielli:2019ivi}, and MATHUSLA \cite{Curtin:2018mvb} could be considered. Still, the heavy spectrum in our parameter space cannot be produced at LHC energies. Similar detectors at a 100~TeV collider might be required. However, for lifetimes of 1~s and large velocities, the probability of detecting sneutrino decays at distances around 100~m is very small. Thus, a long-lived sneutrino in this model is unlikely to be detected by collider experiments.

Finally, it is clear that indirect searches cannot constrain the spectrum in our model. Taking into account that we do not have new sources of flavour, contributions to low-energy observables would be the same in a Constrained MSSM, with Minimal flavour Violation. In this models, processes as $(g-2)_\mu$ or BR($b \to s \gamma$) can not explore multi-TeV sfermion masses \cite{Bertolini:1986tg,Moroi:1995yh,Martin:2001st,Bartl:2001wc}.

In summary, at present is is not possible to explore the spectrum of the model and we must probably wait for a 100 TeV collider to test this model experimentally.

\section{Conclusions/Summary}
In this work, we explore the conditions necessary to obtain a viable asymmetric dark matter  candidate within a supersymmetric extension of the Standard Model. Starting from a supersymmetric leptogenesis scenario, the lepton asymmetry generated by the seesaw sector is redistributed through rapid interactions among particles and sparticles in thermal equilibrium. As a result, part of the lepton asymmetry is carried by the supersymmetric partners of SM particles. In the Minimal Supersymmetric Standard Model, this sparticle asymmetry is efficiently transferred to the SM by sparticle-sparticle annihilations proportional to gaugino Majorana masses and higgsino mass. Since these processes violate $R$-charge, imposing an $R$-symmetry makes it possible to preserve an asymmetry in the (next-to-)lightest supersymmetric particle. However, as we discuss at the end of Section \ref{sec:Cosmo}, with an exact $R$-symmetry the (N)LSP asymmetry would be unrelated to the lepton asymmetry generated by sterile neutrino decays.

To address this limitation, we consider the case in which $R$-symmetry is only approximate. We choose the $R$-breaking terms so that at high temperatures $R$-breaking interactions mediate processes in chemical equilibrium, but consider couplings small enough such that the transitions that transfer the NLSP asymmetry to the SM freeze out once (or before) the temperature reaches the sparticle mass scale. Under these conditions, a sparticle asymmetry is preserved and is related to the lepton asymmetry in a way that can be calculated from the chemical equilibrium conditions. The details on the specific model that we consider are given in Section \ref{sec:Models}, while the cosmology and the evolution of the asymmetry is discussed in Section \ref{sec:detcosmo}. Since the sparticle asymmetry is comparable to the lepton (and therefore baryon) asymmetry, the DM candidate must have a mass close to the proton mass. Since small masses for SM superpartners are clearly excluded, we consider the gravitino as the LSP, as it interacts with the SM particles only via gravitational interactions, making it a viable LSP and DM candidate even for masses $\sim 10$ GeV. Hence, the asymmetric NLSP particle eventually decay into gravitinos, early enough to not disrupt the BBN predictions for light element abundances. We find that NLSP leptons with masses $\gtrsim$ 2 TeV can avoid the BBN bounds (see the end of Section \ref{sec:detcosmo}). The NLSP asymmetry that is transfered to the gravitino abundance can be calculated from the measured baryon asymmetry, and the gravitino mass is predicted once we assume its number density matches the observed DM abundance. Depending on the specific NLSP scenarios, we find that the gravitino should have a mass that ranges from $5$ to $10$ GeV in order to match the observations. Higher gravitino masses may also be viable if $R$-symmetry breaking couplings partially reduce the NLSP asymmetry, making it smaller relative to the baryon asymmetry.

Finally, in Section \ref{sec:pheno} we discuss the spectrum and phenomenology of the model we introduced. We implement the model with \texttt{SARAH-4.15.2} and \texttt{SPheno-4.0.5}, which we use to compute the spectrum. We find regions of parameter space that avoids the appeareance of tachyon states while reproducing the correct SM spectrum. The requirement of NLSP decay before BBN constrains the sparticle spectrum to be in the multi-TeV region. This implies that the model can not be directly tested at the LHC, but could be studied at a future 100 TeV collider. 
\medskip

\paragraph{Acknowledgments.}

We would like to especially thank Sacha Davidson for her collaboration in the early stages of this work and her invaluable discussions. We also thank Werner Porod for help with SPheno.
We acknowledge financial support from the Spanish  Grant PID2023-151418NB-I00 funded by MCIU/AEI/10.13039/501100011033/ FEDER, UE  and from Generalitat Valenciana projects CIPROM/2021/054 and CIPROM/2022/66. DQ acknowledges support from Generalitat Valenciana CIDEGENT/2019/024 and CIESGT2024-021.


\newpage
\appendix

\section{Supersymmetric spectrum}
\label{ap:superspectr}
In this appendix, we discuss the main features of the Supersymmetric spectrum starting from a model defined at high-energies. 

First, we reproduce here the beta functions in the model at two loops for sfermions and gauginos in the limit of zero Yukawa couplings and keeping only leading order terms in $g_1^2$,
\begin{align} 
\label{eq:betafunc}
\beta_{m_{\tilde q}^2} & =  
\frac{1}{5} g_1^2 {\bf 1} \Big(\Sigma + \Delta m^2\Big) 
+\frac{2}{15} \frac{1}{16\pi^2}{\bf 1} \Big(45 g_{2}^{4} \sigma_{2,2}  + 80 g_{3}^{4} \sigma_{2,3} + \frac{1}{10} g_1^2 \sigma_{3,1} \Big)\\ 
\beta_{m_{\tilde l}^2} & =  
 -\frac{3}{5} g_1^2 {\bf 1}  \Big(\Sigma + \Delta m^2\Big)  +\frac{1}{16\pi^2}{\bf 1} \Big(6 g_{2}^{4} \sigma_{2,2}  - \frac{1}{25} g_1^2  \sigma_{3,1} \Big)
 \nonumber \\
 \beta_{m_{\tilde d}^2} & =  
 \frac{2}{5} g_1^2 {\bf 1}  \Big(\Sigma + \Delta m^2\Big) +\frac{8}{15} \frac{1}{16\pi^2}{\bf 1} \Big(20 g_{3}^{4} \sigma_{2,3}  +\frac{1}{20} g_1^2  \sigma_{3,1} \Big)\nonumber \\ 
\beta_{m_{\tilde u}^2} & =  
 - \frac{4}{5} g_1^2 {\bf 1}   \Big(\Sigma + \Delta m^2\Big)+\frac{16}{15} \frac{1}{16\pi^2}{\bf 1} \Big(10 g_{3}^{4} \sigma_{2,3}  - \frac{1}{20} g_1^2  \sigma_{3,1} \Big) \nonumber \\
\beta_{m_{\tilde e}^2} & =  
\frac{6}{5} g_1^2 {\bf 1}  \Big(\Sigma + \Delta m^2\Big) +\frac{1}{16\pi^2}{\bf 1}  \frac{2}{25} g_1^2 \sigma_{3,1} \nonumber \\
\beta_{M^{B}_D} & =  
\frac{2}{5} M^{B}_D \Big(18 g_{1}^{2}  + 5 |\lambda_D|^2  + 5 |\lambda_U|^2 \Big)+\frac{1}{16\pi^2}\Big(
-4 M^{B}_D \Big( |\lambda_D|^4 + |\lambda_U|^4 \Big) +6 M^{B}_D |\lambda_D|^2 \Big( g_{2}^{2}  - |\Lambda_D|^2 \Big)\nonumber \\ 
 &+6 M^{B}_D |\lambda_U|^2 \Big( g_{2}^{2} - |\Lambda_U|^2 \Big) -\frac{1}{5} g_{1}^{2} M^{B}_D \Big( -36 g_{2}^{2}  -\frac{208}{5} g_{1}^{2}  -88 g_{3}^{2}  + 7 \Big( |\Lambda_D|^2  + |\Lambda_U|^2 \Big) \Big) \Big) \Big) \nonumber \\
 \beta_{M^{W}_D} & =  
M^{W}_D \Big(|\Lambda_D|^2 + |\Lambda_U|^2\Big) + \frac{1}{16\pi^2} \Big(
 M^{W}_D \Big(\frac{12}{5} g_{1}^{2} g_{2}^{2} + 88 g_{2}^{4} +24 g_{2}^{2} g_{3}^{2} + \Big(\frac{3}{5} g_{1}^{2} - 8 g_{2}^{2} \Big) \Big( |\Lambda_D|^2 + |\Lambda_U|^2 \Big) \nonumber \\ 
 &-3 \Big(|\Lambda_D|^4 + |\Lambda_U|^4 \Big) -2|\lambda_D|^2 \Big(|\Lambda_D|^2  + g_{2}^{2}\Big)-2 |\lambda_U|^2 \Big(|\Lambda_U|^2  + g_{2}^{2}\Big) \Big) \Big) \nonumber \\ 
 \beta_{M^{O}_D} & =  
-6 g_{3}^{2} M^{O}_D + \frac{1}{16\pi^2}\frac{1}{5} g_{3}^{2} M^{O}_D \Big(11 g_{1}^{2}   + 45 g_{2}^{2}  + 520 g_{3}^{2} \Big) \nonumber
 \end{align}
where, 
\begin{align} \Sigma & = -2 \mbox{Tr}\Big({m_{\tilde u}^2}\Big)  - \mbox{Tr}\Big({m_{\tilde l}^2}\Big) + \mbox{Tr}\Big({m_{\tilde d}^2}\Big) + \mbox{Tr}\Big({m_{\tilde e}^2}\Big) + \mbox{Tr}\Big({m_{\tilde q}^2}\Big)\\ 
 \Delta m^2 &=- m_{H_d}^2  - m_{R_u}^2  + m_{H_u}^2 + m_{R_d}^2 \nonumber \\
\sigma_{3,1} & = 9 \Big( g_1^2 + 5 g_{2}^{2} \Big) \Delta m^2 
 +30 \Big(- m_{R_d}^2  + m_{H_d}^2\Big)|\lambda_D|^2 -30 \Big(- m_{R_u}^2  + m_{H_u}^2\Big)|\lambda_U|^2 \nonumber \\ &+45 \Big( m_{H_d}^2 - m_{R_d}^2 \Big) |\Lambda_D|^2 -45 \Big( m_{H_u}^2 + m_{R_u}^2 \Big) |\Lambda_U|^2 + g_{1}^{2} \Big( 4 \mbox{Tr}\Big({m_{\tilde d}^2}\Big) + 36 \mbox{Tr}\Big({m_{\tilde e}^2}\Big) - 9 \mbox{Tr}\Big({m_{\tilde l}^2}\Big) \Big) \nonumber \\ 
 &  +80 g_{3}^{2} \Big( \mbox{Tr}\Big({m_{\tilde d}^2}\Big)  +\mbox{Tr}\Big({m_{\tilde q}^2}\Big) - 2 \mbox{Tr}\Big({m_{\tilde u}^2}\Big) \Big) -45 g_{2}^{2} \Big(\mbox{Tr}\Big({m_{\tilde l}^2}\Big) +\mbox{Tr}\Big({m_{\tilde q}^2}\Big) \Big) \nonumber\\ 
\sigma_{2,2} & = \frac{1}{2} \Big(3 \mbox{Tr}\Big({m_{\tilde q}^2}\Big)  + 4 m_T^2  + m_{H_d}^2 + m_{H_u}^2 + m_{R_d}^2 + m_{R_u}^2 + \mbox{Tr}\Big({m_{\tilde l}^2}\Big)\Big)\nonumber\\ 
\sigma_{2,3} & = \frac{1}{2} \Big(2 \mbox{Tr}\Big({m_{\tilde q}^2}\Big)  + 6 m_O^2  + \mbox{Tr}\Big({m_{\tilde d}^2}\Big) + \mbox{Tr}\Big({m_{\tilde u}^2}\Big)\Big)\nonumber
\end{align} 
we have to remember that for third generations sfermions, as the stau, the effects of Yukawa couplings reduce further the mass at the electroweak scale from an initial value at high scales.

As explained in the text, the Higgs sector is critical due to the possible presence of tachyons and the need to reproduce the observed Higgs mass.  We follow the analysis of \cite{Diessner:2015yna,kotlarski2016analysisrsymmetricsupersymmetricmodels} including  gluino-sgluon contributions at two loops. In particular, sgluon contributions are positive and rise with the Dirac gluino mass. They can increase the Higgs mass several GeV. Large sgluon and gluino also impacts squarks masses, making them heavier but it does not change significantly other particles masses. On the other hand, we have to control their finite contribution to the pseudoscalar gluon, $\sigma_g$, to avoid it becoming a tachyon. For this we need, $-4M^{D\, 2}_O<B_O<m^2_0$ since
$
m_{\sigma_O}^2=m_0^2-B_O$, and $m_{\phi_O}^2=4M^{D\, 2}_O+B_O
$.

The most dangerous mass matrix for the presence of tachyons is the pseudoscalar mass. We can use a simplified mass matrix at tree-level, in the limit $|\lambda_{U,D}|, |\tilde \lambda_{U,D}|,|\Lambda_{U,D}|\sim {\cal O} (0.1)$, $B_S=B_T=0$, $\tan\beta\gtrsim 10$. In these conditions, and ignoring small terms, the Higgs mass matrices can be expressed in block diagonal form. 
\begin{eqnarray}
    A &=\text{diag}\Big (A_{2\times 2},B_{2\times 2},m_{S_1}^2+\mu_S^2+\mu_{S_2}^2, m_{R_u}^2 + \mu_U^2, m_{R_d}^2 + \mu_D^2,\\
    & M_D^W\left[\frac{\tan^2\beta+1}{\tan^2\beta+1}\frac{g_2}{2} \frac{v^2}{v_T} -4M_D^{W}\right]+\frac{\tan^2\beta}{\tan^2\beta+1}\frac{v^2}{2v_T}\Lambda_u\mu_U\Big )
\end{eqnarray}
where $A_{2\times 2}$ and $B_{2\times 2}$ are
\begin{eqnarray}
    A_{2\times 2} &=&\begin{bmatrix}
        B_\mu\tan\beta & B_\mu\\
        B_\mu & \frac{B_\mu}{\tan\beta}
    \end{bmatrix} \\
    B_{2\times 2} &=&\begin{bmatrix}
        -M_D^B\left[\frac{v^2
   \left(\tan^2\beta-1\right)}{2 v_S
   \left(\tan^2\beta +1\right)}g_1+4M_D^B\right]-\frac{\sqrt{2}
   v^2
   \tan ^2\beta}{2 v_S
   \left(\tan^2\beta+1\right)}\lambda_u \mu_u  & \mu_S \mu_{S_2}\\
        \mu_S \mu_{S_2} & m_{S_1}^2+\mu_{S_2}^2
    \end{bmatrix} \quad 
\end{eqnarray}
By requiring them to have positive eigenvalues we can get 3 approximate condintions from the pseudoscalar matrix:
\begin{align*}
    B_\mu &< 0\\
    \lambda_{U} &<-\frac{\tan^2\beta+1}{\tan^2\beta}\frac{g_1}{\sqrt{2}}\frac{M_D^B}{\mu_U} \simeq - \frac{g_1}{\sqrt{2}}\frac{M_D^B}{\mu_U}\\
    \Lambda_{U} &<-\frac{\tan^2\beta+1}{\tan^2\beta}\frac{g_2}{\sqrt{2}}\frac{M_D^W}{\mu_U} \simeq  -\frac{g_2}{\sqrt{2}}\frac{M_D^W}{\mu_U}
\end{align*}

 Additional conditions are found in the requirement of a slepton NLSP. Under the previous assumptions, and ignoring left-right mixing, selectron and sneutrino mass matrices are
\begin{align}
    m_{\tilde E}^2\simeq\begin{bmatrix}
        m_{\tilde l}^2 & 0\\
        0 & m_{\tilde e}^2
    \end{bmatrix}, \quad 
    m_{\tilde \nu}^2\simeq m_{\tilde l}^2
\end{align}\label{mat:simplified_mass_matrix}
where all the blocks are $3\times 3$. Those matrices at GUT scale are $m_0$ and they run to EW scale. Therefore, we need to examine the conditions in the renormalization group equations (RGEs) to ensure the desired NLSP. Now, we will focus on the one-loop RGEs of the sleptons, although two-loop RGEs were used to obtain the spectrum. Adding the $\tau$ Yukawa coupling to the RGEs, we have,
 \begin{align}
\beta_{m_{\tilde \tau_L}^2}^{(1)} & =  
2 {Y_{\tau}^2} \Big( m_{H_d}^2  + m_{\tilde e,3}^2 + m_{\tilde l,3}^2  \Big) - \frac{3}{5} g_1^2 \Big( \Sigma + \Delta m^2\Big)  \\ 
\beta_{m_{\tilde \tau_R}^2}^{(1)} & =  
4 {Y_\tau^2} \Big( m_{H_d}^2 + m_{\tilde l,3}^2 + m_{\tilde e,3}^2 \Big) +  \frac{6}{5} g_1^2 \Big( \Sigma + \Delta m^2\Big)
\end{align}

If we want the NLSP to be a right-handed slepton, we need to satisfy the condition $\beta_{m_{\tilde \tau_R}^2} > \beta_{m_{\tilde \tau_L}^2}$ and also $\beta_{m_{\tilde \tau_R}^2} > 0$. Then we get
$$
\Sigma-\Delta m^2-\frac{5}{3}\frac{|Y_\tau|^2}{g_1^2}\delta m^2<2 ~m_{R_u}^2<\Sigma-\Delta m^2+\frac{5}{9}\frac{|Y_\tau|^2}{g_1^2}\delta m^2
$$
where $\delta m^2 = m_{\tilde e}^2 + m_{H_d}^2 + m_{\tilde l}^2$.

On the other hand, if we want the NLSP to be a left-handed slepton,  the requirement changes to $\beta_{m_{\tilde \tau_R}^2} < \beta_{m_{\tilde \tau_L}^2}$ with $\beta_{m_{\tilde \tau_L}^2} > 0$, the we get
$$
\Sigma-\Delta m^2+\frac{5}{9}\frac{|Y_\tau|^2}{g_1^2}\delta m^2< 2 ~m_{R_u}^2<\Sigma-\Delta m^2+\frac{5}{3}\frac{|Y_\tau|^2}{g_1^2}\delta m^2
$$
In this case, due to the different sign of the $SU(2)$ D-term contributions to the mass, the sneutrino is naturally lighter than the left-handed selectron and therefore the NLSP. 

\section{Chemical equilibrium equations}\label{app:chemeqs}
In this appendix, we provide additional details on the chemical equilibrium equations for temperature above the SUSY breaking scale. For temperature below the sparticles mass scale, the chemical equilibrium conditions are specified in the text, see Eqs (\ref{eq:chempotlowenergy1}) to (\ref{eq:chempotlowenergy2}). 

Before sparticles decay into the NLSP, the full set of chemical potentials specified in the text in Section \ref{sec:Cosmo}, can be determined from the equilibrium conditions and conserved quantum numbers. We consider temperatures $T\lesssim 10^2-10^3$ TeV, such that all the SM Yukawa interactions are in equilibrium. This means that a process in equilibrium if the corresponding rate is $\Gamma\gtrsim \sqrt{g_\star}\ T^2/M_{\rm pl}\sim 10^{-9}-10^{-11}$ TeV. For example, in a $2\to 2$ scattering for relativistic particles, this implies that, since $\Gamma= n\langle \sigma v\rangle\sim 2T^3/\pi^2 \langle \sigma v\rangle$, the cross-section should be $\langle \sigma v\rangle\gtrsim 10^{-18}-10^{-17} $ ${\rm TeV}^2$.

When the Dirac masses of the gauginos start to become relevant, the particles and sparticles chemical potential could differ because the gauginos can have non-zero asymmetries (while the corresponding gauge bosons have vanishing chemical potentials). Hypercharge gauge interactions in equilibrium fix the particles and sparticles chemical potential differences to be
\begin{equation}
      \mu_{\tilde{f}}-\mu_f=\mu_{\tilde{B}} \label{eq:spartpart}
\end{equation}
for any left-handed particle/sparticle $f, \tilde{f}$ with hypercharge. Since quark doublets have all gauge interactions, it follows that gauginos have the same chemical potential
\begin{equation}
    \mu_\lambda\equiv\mu_{\tilde{B}}=\mu_{\tilde{W}}=\mu_{\tilde{g}}
\end{equation} 
where $\mu_{\tilde{W}},\  \mu_{\tilde{g}}$ is the chemical potential shared by all component of the wino and Gluino gauge multiplets. Accounting for the different $R$ charge of left-handed and right-handed particles, we thus find for all particles/sparticles but the singlets
\begin{equation}
	 \mu_{\tilde{f}_R}-\mu_{f_R}=-\mu_\lambda \qquad \mu_{\tilde{f}_L}-\mu_{f_L}=\mu_\lambda \label{eq:binoeq}
\end{equation}
The $R$ symmetric $\mu_{U,D}$ term (to not be confused with the up and down chemical potentials $\mu_{u,d}$) correspond to Dirac masses for the higgsinos, which in turn lead to the following equalities for the chemical potentials (this can be for instance seen by considering the scattering $h_{u,d} X\leftrightarrow r^*_{u,d} X$ with a $\mu_{U,D}$ insertion in equilibrium )
\begin{align}
	\mu_{h_u}+\mu_{r_u}=0 \nonumber \\
	\mu_{h_d}+\mu_{r_d}=0 
\end{align}
One can similarly argue that the adjoint fermions participating in the gauginos Dirac masses have opposite chemical potentials with respect to the gaugino partner
\begin{equation}
   \mu_{\tilde{B}}=-\mu_{\tilde{S}}\qquad \mu_{\tilde{W}}=-\mu_{\tilde{T}}\qquad \mu_{\tilde{g}}=-\mu_{\tilde{O}}
\end{equation}
Yukawa interactions with quarks in equilibrium impose the following:
\begin{eqnarray}
	\mu_{H_u}-\mu_q+\mu_u=0\nonumber\\
	\mu_{H_d}-\mu_q+\mu_d=0 \label{eq:Yukq}
\end{eqnarray}
while the QCD sphalerons impose (the octet fermion and the gluino contributions cancel)
\begin{eqnarray}	
     2\mu_q-\mu_u-\mu_d=0
\end{eqnarray}
so that summing the conditions in Eq.~(\ref{eq:Yukq}) and assuming the strong sphalerons equilibrium we get
\begin{equation}
	\mu_{H_u}+\mu_{H_d}=0\quad \to \quad \mu_H\equiv \mu_{H_d}=-\mu_{H_u}
\end{equation}
Hypercharge neutrality would instead yield
\begin{align}
	-(\mu_{\ell_\alpha}+2\mu_{\tilde{\ell}_\alpha})-(\mu_{e_\alpha}+2\mu_{\tilde{e}_\alpha})+3(\mu_q+2\mu_{\tilde{q}})+6(\mu_u+2\mu_{\tilde{u}})\nonumber\\
 -3(\mu_d+2\mu_{\tilde{d}})+2(2\mu_H+\mu_{R_u}-\mu_{R_d})+
	+(\mu_{h_u}-\mu_{h_d}+\mu_{r_d}-\mu_{r_u})=0
\end{align}
The superpotential term $\lambda_1 S_1 H_u H_d$ leads to the following Lagrangian
\begin{equation}
	\mathcal{L}\supset \lambda_1 (\tilde{S}_1 h_u) H_d + \lambda_1 S_1(h_u h_d)+{\rm h.c}
\end{equation}
If these interactions are in equilibrium, then the chemical potentials must satisfy 
\begin{align}
	\mu_{{S}_1} + \mu_{h_u}+\mu_{h_d}=0 \qquad \mu_{\tilde{S}_1}+\mu_{h_u}+\mu_{H_d}=0
\end{align}
A sufficiently large $R$-breaking term $B_{\kappa} S_1^3$ ($B_\kappa \sim m_{\rm SUSY}$), imposes that the $S_1$ scalar asymmetry vanishes $\mu_{S_1}=0$, because it can be mediate processes like $S_1+X\to 2S_1+X$. This also implies
\begin{equation}
	 \mu_{h_u}+\mu_{h_d}=0
\end{equation}
The above equation is sufficient to find that, for all non-singlet superfield,   sparticles and particles share the chemical potential 
\begin{align}
    0=\mu_{H_u}+\mu_{H_d}-(\mu_{h_u}+\mu_{h_d})=(\mu_{H_u}-\mu_{h_u})+(\mu_{H_d}-\mu_{h_d})=2\mu_\lambda
\end{align}
where in the last equality we used Eq.~(\ref{eq:binoeq}).  
The superpotential Yukawas between the scalar $S, S_2$ and $r_u h_u$ or $r_d h_d$ also impose the vanishing asymmetries $\mu_S=\mu_{S_2}=0$, while the fermion $\tilde{S}_2$ has a Dirac mass with $\tilde{S}_1$, so that $\mu_{\tilde{S}_1}=-\mu_{\tilde{S}_2}$.
Finally, the electro-weak sphalerons give
\begin{eqnarray}
	9\mu_q+\sum_{\alpha}\mu_{\ell_\alpha}=0,
\end{eqnarray}
where we have used that the extra $SU(2)$ charged particles are "vector-like" and therefore sum to zero in the anomaly.  
The initial lepton asymmetries can be written in terms of the chemical potentials
\begin{eqnarray}
	Y_{B/3-L_\alpha}\equiv Y_{\Delta_\alpha}=\frac{ T^2}{6s}[6(\mu_q+2\mu_{\tilde{q}})+3(\mu_u+2\mu_{\tilde{u}}+\mu_d+2\mu_{\tilde{d}})-2 (\mu_{\ell_\alpha}+2\mu_{\tilde{\ell}_\alpha})-(\mu_{e_\alpha}+2\mu_{\tilde{e}_\alpha})]
\end{eqnarray}
where we have normalized the $B/3-L_\alpha$ number densities to the entropy density $s$ to have the conserved number in the expanding universe. The system of chemical potential equations is solved in terms of $Y_{\Delta_\alpha}$ and the resulting sparticles asymmetry is summarised in the following matrix
\begin{equation}
    \begin{pmatrix}
        \Delta \hat{Y}_{\tilde\ell_e}\\[0.1cm]
        \Delta \hat{Y}_{\tilde\ell_\mu}\\[0.1cm]
        \Delta \hat{Y}_{\tilde\ell_\tau}\\[0.1cm]
        \Delta \hat{Y}_{\tilde e}\\[0.1cm]
        \Delta \hat{Y}_{\tilde \mu}\\[0.1cm]
        \Delta \hat{Y}_{\tilde\tau}\\[0.1cm]
        \Delta \hat{Y}_{\tilde q}\\[0.1cm]
        \Delta \hat{Y}_{\tilde u}\\[0.1cm]
        \Delta \hat{Y}_{\tilde d}\\[0.1cm]
        \Delta \hat{Y}_{h_u}\\[0.1cm]
        \Delta \hat{Y}_{\tilde S_1}\\[0.1cm]
    \end{pmatrix}=\begin{pmatrix}
        -\frac{1300}{3051} & \frac{56}{3051} & \frac{56}{3051} \\[0.1cm]
        \frac{56}{3051} & -\frac{1300}{3051} & \frac{56}{3051} \\[0.1cm]
        \frac{56}{3051} & \frac{56}{3051} & -\frac{1300}{3051} \\[0.1cm]
        -\frac{470}{3051} & \frac{208}{3051} & \frac{208}{3051} \\[0.1cm]
        \frac{208}{3051} & -\frac{470}{3051} & \frac{208}{3051} \\[0.1cm]
        \frac{208}{3051} & \frac{208}{3051} & -\frac{470}{3051} \\[0.1cm]
        \frac{44}{339} & \frac{44}{339} & \frac{44}{339} \\[0.1cm]
        -\frac{38}{339} & -\frac{38}{339} & -\frac{38}{339} \\[0.1cm]
        \frac{82}{339} & \frac{82}{339} & \frac{82}{339} \\[0.1cm]
        -\frac{10}{339} & -\frac{10}{339} & -\frac{10}{339} \\[0.1cm]
        \frac{20}{339} & \frac{20}{339} & \frac{20}{339} \\[0.1cm]
    \end{pmatrix}\times \begin{pmatrix}
        Y_{\Delta_e} \\ 
        Y_{\Delta_\mu} \\
        Y_{\Delta_\tau}
    \end{pmatrix}
\end{equation}


\bibliographystyle{JHEP} 
\bibliography{References}

\providecommand{\href}[2]{#2}\begingroup\raggedright\begin{thebibliography}{10}

\bibitem{Zwicky:1933gu}
F.~Zwicky, {\it {Die Rotverschiebung von extragalaktischen Nebeln}},  {\em Helv. Phys. Acta} {\bf 6} (1933) 110--127.

\bibitem{Rubin:1970zza}
V.~C. Rubin and W.~K. Ford, Jr., {\it {Rotation of the Andromeda Nebula from a Spectroscopic Survey of Emission Regions}},  {\em Astrophys. J.} {\bf 159} (1970) 379--403.

\bibitem{Bertone:2004pz}
G.~Bertone, D.~Hooper, and J.~Silk, {\it {Particle dark matter: Evidence, candidates and constraints}},  {\em Phys. Rept.} {\bf 405} (2005) 279--390, [\href{http://arxiv.org/abs/hep-ph/0404175}{{\tt hep-ph/0404175}}].

\bibitem{Jungman:1995df}
G.~Jungman, M.~Kamionkowski, and K.~Griest, {\it {Supersymmetric dark matter}},  {\em Phys. Rept.} {\bf 267} (1996) 195--373, [\href{http://arxiv.org/abs/hep-ph/9506380}{{\tt hep-ph/9506380}}].

\bibitem{Cirelli:2024ssz}
M.~Cirelli, A.~Strumia, and J.~Zupan, {\it {Dark Matter}},  \href{http://arxiv.org/abs/2406.01705}{{\tt arXiv:2406.01705}}.

\bibitem{Planck:2018vyg}
{\bf Planck} Collaboration, N.~Aghanim et~al., {\it {Planck 2018 results. VI. Cosmological parameters}},  {\em Astron. Astrophys.} {\bf 641} (2020) A6, [\href{http://arxiv.org/abs/1807.06209}{{\tt arXiv:1807.06209}}]. [Erratum: Astron.Astrophys. 652, C4 (2021)].

\bibitem{Gavela:1993ts}
M.~B. Gavela, P.~Hernandez, J.~Orloff, and O.~Pene, {\it {Standard model CP violation and baryon asymmetry}},  {\em Mod. Phys. Lett. A} {\bf 9} (1994) 795--810, [\href{http://arxiv.org/abs/hep-ph/9312215}{{\tt hep-ph/9312215}}].

\bibitem{Huet:1994jb}
P.~Huet and E.~Sather, {\it {Electroweak baryogenesis and standard model CP violation}},  {\em Phys. Rev. D} {\bf 51} (1995) 379--394, [\href{http://arxiv.org/abs/hep-ph/9404302}{{\tt hep-ph/9404302}}].

\bibitem{Kaplan:1991ah}
D.~B. Kaplan, {\it {A Single explanation for both the baryon and dark matter densities}},  {\em Phys. Rev. Lett.} {\bf 68} (1992) 741--743.

\bibitem{Nussinov:1985xr}
S.~Nussinov, {\it {TECHNOCOSMOLOGY: COULD A TECHNIBARYON EXCESS PROVIDE A 'NATURAL' MISSING MASS CANDIDATE?}},  {\em Phys. Lett. B} {\bf 165} (1985) 55--58.

\bibitem{Nardi:2008ix}
E.~Nardi, F.~Sannino, and A.~Strumia, {\it {Decaying Dark Matter can explain the e+- excesses}},  {\em JCAP} {\bf 01} (2009) 043, [\href{http://arxiv.org/abs/0811.4153}{{\tt arXiv:0811.4153}}].

\bibitem{Kaplan:2009ag}
D.~E. Kaplan, M.~A. Luty, and K.~M. Zurek, {\it {Asymmetric Dark Matter}},  {\em Phys. Rev. D} {\bf 79} (2009) 115016, [\href{http://arxiv.org/abs/0901.4117}{{\tt arXiv:0901.4117}}].

\bibitem{Petraki:2013wwa}
K.~Petraki and R.~R. Volkas, {\it {Review of asymmetric dark matter}},  {\em Int. J. Mod. Phys. A} {\bf 28} (2013) 1330028, [\href{http://arxiv.org/abs/1305.4939}{{\tt arXiv:1305.4939}}].

\bibitem{Zurek:2013wia}
K.~M. Zurek, {\it {Asymmetric Dark Matter: Theories, Signatures, and Constraints}},  {\em Phys. Rept.} {\bf 537} (2014) 91--121, [\href{http://arxiv.org/abs/1308.0338}{{\tt arXiv:1308.0338}}].

\bibitem{Shelton:2010ta}
J.~Shelton and K.~M. Zurek, {\it {Darkogenesis: A baryon asymmetry from the dark matter sector}},  {\em Phys. Rev. D} {\bf 82} (2010) 123512, [\href{http://arxiv.org/abs/1008.1997}{{\tt arXiv:1008.1997}}].

\bibitem{Haba:2010bm}
N.~Haba and S.~Matsumoto, {\it {Baryogenesis from Dark Sector}},  {\em Prog. Theor. Phys.} {\bf 125} (2011) 1311--1316, [\href{http://arxiv.org/abs/1008.2487}{{\tt arXiv:1008.2487}}].

\bibitem{Frandsen:2011kt}
M.~T. Frandsen, S.~Sarkar, and K.~Schmidt-Hoberg, {\it {Light asymmetric dark matter from new strong dynamics}},  {\em Phys. Rev. D} {\bf 84} (2011) 051703, [\href{http://arxiv.org/abs/1103.4350}{{\tt arXiv:1103.4350}}].

\bibitem{Haber:1984rc}
H.~E. Haber and G.~L. Kane, {\it {The Search for Supersymmetry: Probing Physics Beyond the Standard Model}},  {\em Phys. Rept.} {\bf 117} (1985) 75--263.

\bibitem{Nilles:1983ge}
H.~P. Nilles, {\it {Supersymmetry, Supergravity and Particle Physics}},  {\em Phys. Rept.} {\bf 110} (1984) 1--162.

\bibitem{Chung:2003fi}
D.~J.~H. Chung, L.~L. Everett, G.~L. Kane, S.~F. King, J.~D. Lykken, and L.-T. Wang, {\it {The Soft supersymmetry breaking Lagrangian: Theory and applications}},  {\em Phys. Rept.} {\bf 407} (2005) 1--203, [\href{http://arxiv.org/abs/hep-ph/0312378}{{\tt hep-ph/0312378}}].

\bibitem{Kang:2011ny}
Z.~Kang and T.~Li, {\it {Asymmetric Origin for Gravitino Relic Density in the Hybrid Gravity-Gauge Mediated Supersymmetry Breaking}},  {\em JHEP} {\bf 10} (2012) 150, [\href{http://arxiv.org/abs/1111.7313}{{\tt arXiv:1111.7313}}].

\bibitem{DEramo:2011dhr}
F.~D'Eramo, L.~Fei, and J.~Thaler, {\it {Dark Matter Assimilation into the Baryon Asymmetry}},  {\em JCAP} {\bf 03} (2012) 010, [\href{http://arxiv.org/abs/1111.5615}{{\tt arXiv:1111.5615}}].

\bibitem{Davidson:2008bu}
S.~Davidson, E.~Nardi, and Y.~Nir, {\it {Leptogenesis}},  {\em Phys. Rept.} {\bf 466} (2008) 105--177, [\href{http://arxiv.org/abs/0802.2962}{{\tt arXiv:0802.2962}}].

\bibitem{Affleck:1984fy}
I.~Affleck and M.~Dine, {\it {A New Mechanism for Baryogenesis}},  {\em Nucl. Phys. B} {\bf 249} (1985) 361--380.

\bibitem{Pilaftsis:1997jf}
A.~Pilaftsis, {\it {CP violation and baryogenesis due to heavy Majorana neutrinos}},  {\em Phys. Rev. D} {\bf 56} (1997) 5431--5451, [\href{http://arxiv.org/abs/hep-ph/9707235}{{\tt hep-ph/9707235}}].

\bibitem{Hambye:2000zs}
T.~Hambye, E.~Ma, and U.~Sarkar, {\it {Leptogenesis from neutralino decay with nonholomorphic R-parity violation}},  {\em Nucl. Phys. B} {\bf 590} (2000) 429--452, [\href{http://arxiv.org/abs/hep-ph/0006173}{{\tt hep-ph/0006173}}].

\bibitem{Hambye:2001eu}
T.~Hambye, {\it {Leptogenesis at the TeV scale}},  {\em Nucl. Phys. B} {\bf 633} (2002) 171--192, [\href{http://arxiv.org/abs/hep-ph/0111089}{{\tt hep-ph/0111089}}].

\bibitem{Moroi:1995fs}
T.~Moroi, {\it {Effects of the gravitino on the inflationary universe}},  other thesis, 3, 1995.

\bibitem{Khlebnikov:1988sr}
S.~Y. Khlebnikov and M.~E. Shaposhnikov, {\it {The Statistical Theory of Anomalous Fermion Number Nonconservation}},  {\em Nucl. Phys. B} {\bf 308} (1988) 885--912.

\bibitem{Harvey:1990qw}
J.~A. Harvey and M.~S. Turner, {\it {Cosmological baryon and lepton number in the presence of electroweak fermion number violation}},  {\em Phys. Rev. D} {\bf 42} (1990) 3344--3349.

\bibitem{Asaka:2000zh}
T.~Asaka, K.~Hamaguchi, and K.~Suzuki, {\it {Cosmological gravitino problem in gauge mediated supersymmetry breaking models}},  {\em Phys. Lett. B} {\bf 490} (2000) 136--146, [\href{http://arxiv.org/abs/hep-ph/0005136}{{\tt hep-ph/0005136}}].

\bibitem{Fayet:1974pd}
P.~Fayet, {\it {Supergauge Invariant Extension of the Higgs Mechanism and a Model for the electron and Its Neutrino}},  {\em Nucl. Phys. B} {\bf 90} (1975) 104--124.

\bibitem{Hall:1990hq}
L.~J. Hall and L.~Randall, {\it {U(1)-R symmetric supersymmetry}},  {\em Nucl. Phys. B} {\bf 352} (1991) 289--308.

\bibitem{Randall:1992cq}
L.~Randall and N.~Rius, {\it {The Minimal U(1)-R symmetric model revisited}},  {\em Phys. Lett. B} {\bf 286} (1992) 299--306.

\bibitem{Nelson:2002ca}
A.~E. Nelson, N.~Rius, V.~Sanz, and M.~Unsal, {\it {The Minimal supersymmetric model without a mu term}},  {\em JHEP} {\bf 08} (2002) 039, [\href{http://arxiv.org/abs/hep-ph/0206102}{{\tt hep-ph/0206102}}].

\bibitem{Fox:2002bu}
P.~J. Fox, A.~E. Nelson, and N.~Weiner, {\it {Dirac gaugino masses and supersoft supersymmetry breaking}},  {\em JHEP} {\bf 08} (2002) 035, [\href{http://arxiv.org/abs/hep-ph/0206096}{{\tt hep-ph/0206096}}].

\bibitem{Abel:2013kha}
S.~Abel and D.~Busbridge, {\it {Mapping Dirac gaugino masses}},  {\em JHEP} {\bf 11} (2013) 098, [\href{http://arxiv.org/abs/1306.6323}{{\tt arXiv:1306.6323}}].

\bibitem{Diessner:2014ksa}
P.~Die\ss{}ner, J.~Kalinowski, W.~Kotlarski, and D.~St\"ockinger, {\it {Higgs boson mass and electroweak observables in the MRSSM}},  {\em JHEP} {\bf 12} (2014) 124, [\href{http://arxiv.org/abs/1410.4791}{{\tt arXiv:1410.4791}}].

\bibitem{abel2011easy}
S.~Abel and M.~Goodsell, {\it Easy dirac gauginos},  {\em Journal of High Energy Physics} {\bf 2011} (2011), no.~6 1--33.

\bibitem{benakli2011dirac}
K.~Benakli, {\it Dirac gauginos: a user manual},  {\em Fortschritte der Physik} {\bf 59} (2011), no.~11-12 1079--1082.

\bibitem{benakli2013dirac}
K.~Benakli, M.~D. Goodsell, and F.~Staub, {\it Dirac gauginos and the 125 gev higgs},  {\em Journal of High Energy Physics} {\bf 2013} (2013), no.~6 1--34.

\bibitem{benakli2017framework}
K.~Benakli, {\it A framework for unified dirac gauginos},  in {\em EPJ Web of Conferences}, vol.~164, p.~01001, EDP Sciences, 2017.

\bibitem{Hisano:2006nn}
J.~Hisano, S.~Matsumoto, M.~Nagai, O.~Saito, and M.~Senami, {\it {Non-perturbative effect on thermal relic abundance of dark matter}},  {\em Phys. Lett. B} {\bf 646} (2007) 34--38, [\href{http://arxiv.org/abs/hep-ph/0610249}{{\tt hep-ph/0610249}}].

\bibitem{Feng:2004zu}
J.~L. Feng, S.-f. Su, and F.~Takayama, {\it {SuperWIMP gravitino dark matter from slepton and sneutrino decays}},  {\em Phys. Rev. D} {\bf 70} (2004) 063514, [\href{http://arxiv.org/abs/hep-ph/0404198}{{\tt hep-ph/0404198}}].

\bibitem{Boubekeur:2010nt}
L.~Boubekeur, K.~Y. Choi, R.~Ruiz~de Austri, and O.~Vives, {\it {The degenerate gravitino scenario}},  {\em JCAP} {\bf 04} (2010) 005, [\href{http://arxiv.org/abs/1002.0340}{{\tt arXiv:1002.0340}}].

\bibitem{Kawasaki:2004yh}
M.~Kawasaki, K.~Kohri, and T.~Moroi, {\it {Hadronic decay of late - decaying particles and Big-Bang Nucleosynthesis}},  {\em Phys. Lett. B} {\bf 625} (2005) 7--12, [\href{http://arxiv.org/abs/astro-ph/0402490}{{\tt astro-ph/0402490}}].

\bibitem{porod2003spheno}
W.~Porod, {\it Spheno, a program for calculating supersymmetric spectra, susy particle decays and susy particle production at e+ e- colliders},  {\em Computer Physics Communications} {\bf 153} (2003), no.~2 275--315.

\bibitem{Porod_2012}
W.~Porod and F.~Staub, {\it Spheno 3.1: extensions including flavour, cp-phases and models beyond the mssm},  {\em Computer Physics Communications} {\bf 183} (Nov., 2012) 2458–2469.

\bibitem{Pierce:1996zz}
D.~M. Pierce, J.~A. Bagger, K.~T. Matchev, and R.-j. Zhang, {\it {Precision corrections in the minimal supersymmetric standard model}},  {\em Nucl. Phys. B} {\bf 491} (1997) 3--67, [\href{http://arxiv.org/abs/hep-ph/9606211}{{\tt hep-ph/9606211}}].

\bibitem{Degrassi:2001yf}
G.~Degrassi, P.~Slavich, and F.~Zwirner, {\it {On the neutral Higgs boson masses in the MSSM for arbitrary stop mixing}},  {\em Nucl. Phys. B} {\bf 611} (2001) 403--422, [\href{http://arxiv.org/abs/hep-ph/0105096}{{\tt hep-ph/0105096}}].

\bibitem{Brignole:2001jy}
A.~Brignole, G.~Degrassi, P.~Slavich, and F.~Zwirner, {\it {On the O(alpha(t)**2) two loop corrections to the neutral Higgs boson masses in the MSSM}},  {\em Nucl. Phys. B} {\bf 631} (2002) 195--218, [\href{http://arxiv.org/abs/hep-ph/0112177}{{\tt hep-ph/0112177}}].

\bibitem{Brignole:2002bz}
A.~Brignole, G.~Degrassi, P.~Slavich, and F.~Zwirner, {\it {On the two loop sbottom corrections to the neutral Higgs boson masses in the MSSM}},  {\em Nucl. Phys. B} {\bf 643} (2002) 79--92, [\href{http://arxiv.org/abs/hep-ph/0206101}{{\tt hep-ph/0206101}}].

\bibitem{Dedes:2002dy}
A.~Dedes and P.~Slavich, {\it {Two loop corrections to radiative electroweak symmetry breaking in the MSSM}},  {\em Nucl. Phys. B} {\bf 657} (2003) 333--354, [\href{http://arxiv.org/abs/hep-ph/0212132}{{\tt hep-ph/0212132}}].

\bibitem{Dedes:2003km}
A.~Dedes, G.~Degrassi, and P.~Slavich, {\it {On the two loop Yukawa corrections to the MSSM Higgs boson masses at large tan beta}},  {\em Nucl. Phys. B} {\bf 672} (2003) 144--162, [\href{http://arxiv.org/abs/hep-ph/0305127}{{\tt hep-ph/0305127}}].

\bibitem{staub2012sarah}
F.~Staub, {\it Sarah},  2012.

\bibitem{Staub_2014}
F.~Staub, {\it Sarah   4: A tool for (not only susy) model builders},  {\em Computer Physics Communications} {\bf 185} (June, 2014) 1773–1790.

\bibitem{goodsell2023active}
M.~D. Goodsell and A.~Joury, {\it Active learning bsm parameter spaces},  {\em The European Physical Journal C} {\bf 83} (2023), no.~4 268.

\bibitem{kotlarski2016analysisrsymmetricsupersymmetricmodels}
W.~Kotlarski, {\it Analysis of the r-symmetric supersymmetric models including quantum corrections},  2016.

\bibitem{Nilles:1982dy}
H.~P. Nilles, M.~Srednicki, and D.~Wyler, {\it {Weak Interaction Breakdown Induced by Supergravity}},  {\em Phys. Lett. B} {\bf 120} (1983) 346.

\bibitem{Alvarez-Gaume:1983drc}
L.~Alvarez-Gaume, J.~Polchinski, and M.~B. Wise, {\it {Minimal Low-Energy Supergravity}},  {\em Nucl. Phys. B} {\bf 221} (1983) 495.

\bibitem{Claudson:1983et}
M.~Claudson, L.~J. Hall, and I.~Hinchliffe, {\it {Low-Energy Supergravity: False Vacua and Vacuous Predictions}},  {\em Nucl. Phys. B} {\bf 228} (1983) 501--528.

\bibitem{Gunion:1987qv}
J.~F. Gunion, H.~E. Haber, and M.~Sher, {\it {Charge / Color Breaking Minima and a-Parameter Bounds in Supersymmetric Models}},  {\em Nucl. Phys. B} {\bf 306} (1988) 1--13.

\bibitem{Casas:1995pd}
J.~A. Casas, A.~Lleyda, and C.~Munoz, {\it {Strong constraints on the parameter space of the MSSM from charge and color breaking minima}},  {\em Nucl. Phys. B} {\bf 471} (1996) 3--58, [\href{http://arxiv.org/abs/hep-ph/9507294}{{\tt hep-ph/9507294}}].

\bibitem{Camargo-Molina:2014pwa}
J.~E. Camargo-Molina, B.~Garbrecht, B.~O'Leary, W.~Porod, and F.~Staub, {\it {Constraining the Natural MSSM through tunneling to color-breaking vacua at zero and non-zero temperature}},  {\em Phys. Lett. B} {\bf 737} (2014) 156--161, [\href{http://arxiv.org/abs/1405.7376}{{\tt arXiv:1405.7376}}].

\bibitem{atlas2017search}
A.~Collaboration et~al., {\it Search for squarks and gluinos in final states with jets and missing transverse momentum using 36 fb$^{-1}$ of $\sqrt s= 13$ tev $pp$ collision data with the atlas detector},  {\em arXiv preprint arXiv:1712.02332} (2017).

\bibitem{chalons2019lhc}
G.~Chalons, M.~D. Goodsell, S.~Kraml, H.~Reyes-Gonz{\'a}lez, and S.~L. Williamson, {\it Lhc limits on gluinos and squarks in the minimal dirac gaugino model},  {\em Journal of High Energy Physics} {\bf 2019} (2019), no.~4 1--28.

\bibitem{GrillidiCortona:2016pyt}
G.~Grilli~di Cortona, E.~Hardy, and A.~J. Powell, {\it {Dirac vs Majorana gauginos at a 100 TeV collider}},  {\em JHEP} {\bf 08} (2016) 014, [\href{http://arxiv.org/abs/1606.07090}{{\tt arXiv:1606.07090}}].

\bibitem{Goodsell:2020lpx}
M.~D. Goodsell, S.~Kraml, H.~Reyes-Gonz\'alez, and S.~L. Williamson, {\it {Constraining Electroweakinos in the Minimal Dirac Gaugino Model}},  {\em SciPost Phys.} {\bf 9} (2020), no.~4 047, [\href{http://arxiv.org/abs/2007.08498}{{\tt arXiv:2007.08498}}].

\bibitem{Carpenter:2021tnq}
L.~M. Carpenter and M.~J. Smylie, {\it {Exploring the phenomenology of weak adjoint scalars in minimal R-symmetric models}},  {\em JHEP} {\bf 02} (2022) 102, [\href{http://arxiv.org/abs/2108.02795}{{\tt arXiv:2108.02795}}].

\bibitem{Feng:2015wqa}
J.~L. Feng, S.~Iwamoto, Y.~Shadmi, and S.~Tarem, {\it {Long-Lived Sleptons at the LHC and a 100 TeV Proton Collider}},  {\em JHEP} {\bf 12} (2015) 166, [\href{http://arxiv.org/abs/1505.02996}{{\tt arXiv:1505.02996}}].

\bibitem{ATLAS:2017oal}
{\bf ATLAS} Collaboration, M.~Aaboud et~al., {\it {Search for long-lived charginos based on a disappearing-track signature in pp collisions at $ \sqrt{s}=13 $ TeV with the ATLAS detector}},  {\em JHEP} {\bf 06} (2018) 022, [\href{http://arxiv.org/abs/1712.02118}{{\tt arXiv:1712.02118}}].

\bibitem{ATLAS:2022rme}
{\bf ATLAS} Collaboration, G.~Aad et~al., {\it {Search for long-lived charginos based on a disappearing-track signature using 136 fb$^{-1}$ of pp collisions at $\sqrt{s}$~=~13~TeV with the ATLAS detector}},  {\em Eur. Phys. J. C} {\bf 82} (2022), no.~7 606, [\href{http://arxiv.org/abs/2201.02472}{{\tt arXiv:2201.02472}}].

\bibitem{CMS:2018rea}
{\bf CMS} Collaboration, A.~M. Sirunyan et~al., {\it {Search for disappearing tracks as a signature of new long-lived particles in proton-proton collisions at $\sqrt{s} =$ 13 TeV}},  {\em JHEP} {\bf 08} (2018) 016, [\href{http://arxiv.org/abs/1804.07321}{{\tt arXiv:1804.07321}}].

\bibitem{CMS:2020atg}
{\bf CMS} Collaboration, A.~M. Sirunyan et~al., {\it {Search for disappearing tracks in proton-proton collisions at $\sqrt{s} =$ 13 TeV}},  {\em Phys. Lett. B} {\bf 806} (2020) 135502, [\href{http://arxiv.org/abs/2004.05153}{{\tt arXiv:2004.05153}}].

\bibitem{Goodsell:2021iwc}
M.~D. Goodsell and L.~Priya, {\it {Long dead winos}},  {\em Eur. Phys. J. C} {\bf 82} (2022), no.~3 235, [\href{http://arxiv.org/abs/2106.08815}{{\tt arXiv:2106.08815}}].

\bibitem{MoEDAL:2014ttp}
{\bf MoEDAL} Collaboration, B.~Acharya et~al., {\it {The Physics Programme Of The MoEDAL Experiment At The LHC}},  {\em Int. J. Mod. Phys. A} {\bf 29} (2014) 1430050, [\href{http://arxiv.org/abs/1405.7662}{{\tt arXiv:1405.7662}}].

\bibitem{Pinfold:2019zwp}
J.~L. Pinfold, {\it {The MoEDAL experiment: a new light on the high-energy frontier}},  {\em Phil. Trans. Roy. Soc. Lond. A} {\bf 377} (2019), no.~2161 20190382.

\bibitem{Aielli:2019ivi}
G.~Aielli et~al., {\it {Expression of interest for the CODEX-b detector}},  {\em Eur. Phys. J. C} {\bf 80} (2020), no.~12 1177, [\href{http://arxiv.org/abs/1911.00481}{{\tt arXiv:1911.00481}}].

\bibitem{Curtin:2018mvb}
D.~Curtin et~al., {\it {Long-Lived Particles at the Energy Frontier: The MATHUSLA Physics Case}},  {\em Rept. Prog. Phys.} {\bf 82} (2019), no.~11 116201, [\href{http://arxiv.org/abs/1806.07396}{{\tt arXiv:1806.07396}}].

\bibitem{Bertolini:1986tg}
S.~Bertolini, F.~Borzumati, and A.~Masiero, {\it {New Constraints on Squark and Gluino Masses from Radiative b Decays}},  {\em Phys. Lett. B} {\bf 192} (1987) 437--440.

\bibitem{Moroi:1995yh}
T.~Moroi, {\it {The Muon anomalous magnetic dipole moment in the minimal supersymmetric standard model}},  {\em Phys. Rev. D} {\bf 53} (1996) 6565--6575, [\href{http://arxiv.org/abs/hep-ph/9512396}{{\tt hep-ph/9512396}}]. [Erratum: Phys.Rev.D 56, 4424 (1997)].

\bibitem{Martin:2001st}
S.~P. Martin and J.~D. Wells, {\it {Muon Anomalous Magnetic Dipole Moment in Supersymmetric Theories}},  {\em Phys. Rev. D} {\bf 64} (2001) 035003, [\href{http://arxiv.org/abs/hep-ph/0103067}{{\tt hep-ph/0103067}}].

\bibitem{Bartl:2001wc}
A.~Bartl, T.~Gajdosik, E.~Lunghi, A.~Masiero, W.~Porod, H.~Stremnitzer, and O.~Vives, {\it {General flavor blind MSSM and CP violation}},  {\em Phys. Rev. D} {\bf 64} (2001) 076009, [\href{http://arxiv.org/abs/hep-ph/0103324}{{\tt hep-ph/0103324}}].

\bibitem{Diessner:2015yna}
P.~Diessner, J.~Kalinowski, W.~Kotlarski, and D.~St\"ockinger, {\it {Two-loop correction to the Higgs boson mass in the MRSSM}},  {\em Adv. High Energy Phys.} {\bf 2015} (2015) 760729, [\href{http://arxiv.org/abs/1504.05386}{{\tt arXiv:1504.05386}}].

\end{thebibliography}\endgroup

\end{document}